# Density of States Weighted Decoherence Probe Formalism for Charge Transport in DNA

Hashem Mohammad[1*] and M.P. Anantram[2].

[1] Department of Electrical Engineering, College of Engineering and Petroleum, Kuwait University, Sabah Al Salem University City, P.O. Box 5969, Safat 13060, Shadadiya.

[2] Department of Electrical and Computer Engineering, University of Washington, Seattle, Washington 98115, United States.

[*] Corresponding author

## Abstract

Nanoscale molecular systems such as DNA require an atomistic quantum treatment to accurately capture their electrical properties, owing to their small dimensions. A central challenge in modeling transport through these systems is the inclusion of phase-breaking scattering. Decoherence-probe methods enable such modeling for large systems, but existing implementations have limitations. Energy-independent scattering rates tend to overly broaden energy levels, yielding an unphysically large density of states (DOS) within energy gaps. Conversely, energy-dependent models may introduce spurious energy levels and transmission peaks and require additional fitting parameters. To address these issues, we use a DOS-weighted decoherence model in which the scattering rate and equivalently, the associated decoherence probe self-energy is proportional to the local DOS. The model iteratively updates the decoherence self-energy and the DOS until self-consistency is achieved. This approach yields energy and spatially dependent scattering rates that avoid spurious energy levels without the excessive broadening of DOS in energy gaps. We also examine the impact of partitioning schemes that prevent artificial pathways for charge transport and discuss how they can be avoided. Overall, the DOS-weighted model provides an improved and more physically grounded framework for simulating charge transport in DNA and potentially other weakly coupled molecular systems.

# 1 Introduction

The field of molecular electronics is rapidly advancing, enabling the development of nanoscale devices and sensors that extend beyond conventional solid-state semiconductors. Nucleic acids are among the prominent candidates for such applications due to their properties of self-assembly, molecular recognition, and charge transport. The integration of these characteristics facilitates bottom-up fabrication strategies for a wide range of nanoscale structures [1,2]. Moreover, the ability to conduct electrical current at measurable ranges allows for use in nanoelectronics. For instance, DNA strands between 8-15 base-pairs, corresponding to 2-5 nm in length, can have conductance within the nS-μS range ($\mu G_0 - mG_0$, where $G_0$ is conductance quantum) [3–6]. DNA conductance can be modulated through various processes, including the intercalation of molecules between base pairs or within the DNA groove [5–8], base substitution (e.g., replacing a native base with graphene quantum dot [9]), chemical modification of nucleotides such as methylation or base-pair mismatch [10–12], metal ion doping [13], and aptamer binding [14]. Potential electronic applications tunnel barriers and quantum wells [15,16] encompass memory and data storage devices [13,17], and biosensors [5,10,12,18,19]. Consequently, elucidating the fundamental mechanisms of charge transport in nanoscale DNA structures is critical for advancing DNA-based molecular electronics.

DNA conductance is commonly measured by connecting a DNA strand to two metallic contacts at its two ends and applying a bias. The reported experimentally observed electrical behavior of DNA ranges from insulating to Ohmic to semiconducting to superconducting, depending on the measurement conditions [3,5,20–24]. However, DNA in a solvent has a bandgap of 3eV or larger as per DFT studies [8,12,14,16,25,26]. The wide variation in experimental conductance values observed is perhaps due to the possibility of having the Fermi energy anywhere between the HOMO and LUMO levels due to factors such as temperature, contact geometry and material, unintentional doping, and possible polaron formation. Additionally, the position of the contacts relative to the DNA bases influences the accessible localized states near to the electrode's Fermi energy, thereby affecting charge transport and conductance [19,27].

Broadly, as an electron travels between the contacts, a variety of physical phenomena should play a role in determining the conductance: reorganization energy and change in atomic coordinates due to DNA-solvent interaction, conventional scattering between vibrational modes and electrons, polaron formation, thermal motion of ions and solvent molecules in regions around the DNA causing fluctuations in the energy levels of the DNA, and electron-electron interaction. These effects break the electron's phase and broaden energy levels in DNA, amongst other physical effects. Approaches based on Marcus theory and polaron hopping are methods that account for some of the above-mentioned physical phenomena. We present a density-of-states-weighted (DOS-weighted) decoherence probe model for charge transport through DNA in this manuscript which partially accounts for the mentioned effects by including decoherence, albeit in a phenomenological manner.

The electronic state is fundamentally a many-body wavefunction, and the loss of coherence occurs by the entanglement of this electronic wavefunction with external degrees of freedom. Our treatment does not explicitly capture either the many-body wavefunction or entanglement with the external degrees of freedom; instead, we employ a single-particle approximation in which each electron exchanges phase with a decoherence probe. Although multi-electron correlation effects are neglected, the model remains tractable for systems containing many thousands of orbitals and reproduces the essential physics of decoherence. While we adopt a single-particle Green's function method with decoherence probes, it is important to note that the modeling of decoherence has been approached using other methodologies. References [28–30] employ coupling to a phonon bath, where decoherence arises from entanglement of the electron and bath degrees of freedom and eventual loss of phase to the bath which is maintained at equilibrium. A different strategy is adopted in reference [31], which incorporates environmental fluctuations directly through a time-dependent electronic Hamiltonian. The decoherence probe approach models scattering by attaching one or more fictitious probes to the DNA bases. The electron may scatter into the probe, lose its phase coherence, and then be reinjected into the DNA to continue traversing the structure. The scattering into the probe is controlled by a single parameter, in this phenomenological model for phase-breaking processes. The scattering rate was initially taken to have an energy-independent value [32–35]. However, this approach produces unphysically large density of states (DOS) within energy gaps of DNA, which excessively broadens the DOS peaks. To address this, we extended the scattering rate definition to be energy dependent, which mitigates the unphysically large DOS in the HOMO-LUMO and other energy gaps [36]. However, this approach introduces spurious peaks within energy gaps and in addition requires an additional fitting parameter when comparing with experimental data.

In this work, we propose a density-of-states-weighted decoherence probe model, in which the decoherence self-energy is proportional to the DOS. The Hamiltonian of DNA obtained from density functional theory (though it could be obtained from the tight binding method also) is used in the equation for the retarded Green's function to calculate the DOS which depends on both position and energy. At each energy, the retarded Green's function and decoherence probe self-energy are solved self-consistently: the DOS and self-energy are iteratively updated until convergence to a prescribed accuracy is achieved. The resulting retarded Green's function is then used to compute the transmission and conductance.

We also investigate the effect of decoherence probe implementation; the decoherence probe can in principle be attached to an orbital, an atom, a base or to a larger substructure. To attach the probes, the DNA molecule must be partitioned into sub-blocks, which involves diagonalizing the sub-Hamiltonian of each partition to obtain the localized molecular orbitals within that sub-block. The probes are then attached to these localized orbitals. This sub-block partitioning can be defined in various ways: probes may be attached to individual atoms, all atoms comprising a nucleotide, or all atoms within a base-pair, or multiple nucleotides. Such partitioning can introduce two issues. First, it can artificially break covalent C-O bonds and introduce spurious peaks in the energy gap. Our density-of-states-weighted model eliminates these spurious peaks, providing a more physical description of charge transport in DNA. Second, the partitioning scheme can create unphysical shortcut pathways: If a partition contains too many bases, electrons may bypass intermediate sites, leading to artificially overestimating transmission and conductance. While this latter issue is inherent to the partitioning scheme, it highlights the need for careful consideration when defining sub-blocks for decoherence probe attachment.

The paper is organized as follows. Section 2 discusses the methodology, and we summarize the difference between the decoherence probe model implementations. In Section 3, we discuss the effect of including the real part of the decoherence probe self-energy on transmission. Section 4 presents the DNA charge transport results, including the converged decoherence rates and scattering times. Finally, Section 5 discusses partitioning, its role in inducing spurious peaks, and the comparison of different partitioning schemes.

# 2 Methods

For a direct comparison with our previous work [36], we model a double-strand B-DNA 3'-CCCGCGCCC-5' using density functional theory (DFT) to generate the Hamiltonian of the DNA and calculate the transmission via the Green's function method, incorporating decoherence probes to account for phase-breaking processes. For the charge transport calculations, we assume the metallic contacts to be attached to the cytosine nucleotides at both the 3' and 5' ends of the strand. We emphasize that this approach does not rely on using the less-than and greater-than Green's function to model decoherence [37] and that only the retarded and advanced Green's functions are required.

## 2.1 Theory and Formalism

The atomic structure of the B-DNA strand is constructed using the Nucleic Acid Builder module in Amber [38]. Based on the generated coordinates, single-point DFT calculations are performed using the B3LYP functionals and the 6-31G(d,p) basis set [39]. In our previous work, we found that B3LYP and CAM-B3LYP give similar transmission profiles [11] though the bandgaps were different. We also found that the 6-31G(d,p) basis set produces energy levels and transmission profile in the HOMO and LUMO bands that are similar to those from larger basis sets 6-311G(d,p), cc-pVDZ, and cc-pVTZ (see Supplementary Information of Refs [36,40]). The effect of the aqueous environment surrounding the DNA is included through the polarizable continuum model (PCM) with the integral equation formalism (IEFPCM). The DFT calculation yields the Fock (F) and Overlap (S) matrices, which we subsequently use as inputs for the charge transport calculations.

We apply a Löwdin transformation to obtain the Hamiltonian ($H_0$) in an orthogonal basis set

$$H_0 = S^{-\frac{1}{2}} F S^{-\frac{1}{2}} \tag{1}$$

The Hamiltonian of the DNA can be expressed as

$$H_0 = H_0^D + H_0^{OD} \tag{2}$$

$$H_0^D = \sum_{\alpha=1}^{b_i} \sum_{i=1}^{N_{DNA}} \epsilon_{\alpha,i} \, c_{\alpha,i}^\dagger \, c_{\alpha,i} \tag{3}$$

$$H_0^{OD} = \sum_{\alpha=1}^{b_i} \sum_{\alpha'=1}^{b_j} \sum_{i,j=1}^{N_{DNA}} t_{\alpha,i\to\alpha',j} \left( c_{\alpha,i}^\dagger \, c_{\alpha',j} + H.c \right) \tag{4}$$

In this expression, $\epsilon_{\alpha,i}$ denotes the on-site energy of orbital $\alpha$ at atom $i$, while $c_{\alpha,i}^\dagger$ and $c_{\alpha,i}$ are the creation and annihilation operators that create and remove an electron in orbital $\alpha$ on atom $i$. The term $t_{\alpha,i\to\alpha',j}$ represents the coupling between orbital $\alpha$ on atom $i$ and orbital $\alpha'$ on atom $j$, with $H.c$ indicating the Hermitian conjugate. The quantity $b_i$ refers to the number of basis functions representing atom $i$, and $N_{DNA}$ is the total number of atoms in the DNA strand.

$H_0^D$ and $H_0^{OD}$ are the diagonal and off-diagonal parts of the Hamiltonian, respectively. The diagonal elements of $H_0$ correspond to the energy levels of individual atomic orbitals, while the off-diagonal elements describe the hopping parameters between them. The full Hamiltonian $H_0$ is a square matrix, with its size determined by the total number of basis functions in the system, given by $\sum_{i=1}^{N_{DNA}} b_i$.

Following our approach in [35] and [36], we partition the DNA into individual nucleotides (each consisting of a backbone and a base) as shown in Figure 1. The $m^{th}$ diagonal block of the Hamiltonian $H_0$ is the sub-Hamiltonian $H_{0m}$ associated with nucleotide $m$, which is partition $m$. This sub-Hamiltonian consists of the Hamiltonian terms representing the interaction between all atomic orbitals of nucleotide $m$. Let $u_m$ be a diagonal sub-matrix composed of the eigenvectors of $H_{0m}$. We then define the matrix $U$ as follows:

$$U = \begin{bmatrix} u_1 & 0 & & \cdots & & 0 \\ 0 & u_2 & & & & \\ & & \ddots & & & \vdots \\ \vdots & & & u_m & & \\ & & & & \ddots & 0 \\ 0 & & \cdots & & 0 & u_N \end{bmatrix} \tag{5}$$

where $N$ denotes the total number of nucleotides in the DNA, and the dimension of $u_m$ is $O_m$, which is defined as:

$$O_m = \sum_{i=1}^{N_m} b_i \tag{6}$$

$O_m$ represents the total number of basis functions used to describe nucleotide $m$, while $N_m$ is the total number of atoms in nucleotide $m$. We then define a unitary transformation,

$$H = U^{\dagger} H_0 U \tag{7}$$

This unitary transformation diagonalizes the diagonal blocks corresponding to each nucleotide $m$, resulting in diagonal matrices for those blocks in the transformed Hamiltonian $H$. The diagonal elements within each block correspond to the eigenvalues of nucleotide $m$. The off-diagonal blocks of $H$ represent the hopping parameters between the molecular orbitals of different nucleotides. Each diagonal block associated with nucleotide $m$ has a size of $O_m$.

The transformed Hamiltonian $H$ expressed in second quantized form is given by:

$$H = H^D + H^{OD} \tag{8}$$

$$H_0^D = \sum_{k=1}^{O_m} \sum_{m=1}^{N} \tilde{\epsilon}_{k,m}\, d_{k,m}^{\dagger}\, d_{k,m} \tag{9}$$

$$H_0^{OD} = \sum_{k=1}^{O_m} \sum_{k'=1}^{O_{m'}} \sum_{\substack{\mathrm{m}=1 \\ m'=1 \\ \mathrm{m} \neq \mathrm{m}'}}^{N} \tilde{t}_{k,m\rightarrow k',m'} \left( d_{k,m}^{\dagger}\, d_{k',m'} + H.c \right) \tag{10}$$

where $\tilde{\epsilon}_{k,m}$ is the $k^{\text{th}}$ molecular orbital energy of the $m^{\text{th}}$ partition/nucleotide, and $\tilde{t}_{k,m\to k',m'}$ is the coupling between the $k^{\text{th}}$ molecular orbital of the $m^{\text{th}}$ nucleotide with the $k'^{\text{th}}$ molecular orbital of the $m'^{\text{th}}$ nucleotide, and $N$ is the total number of nucleotides. $d^{\dagger}_{k,m}$ and $d_{k,m}$ represent the creation and annihilation operators that create and remove an electron at molecular orbital $k$ in nucleotide $m$.

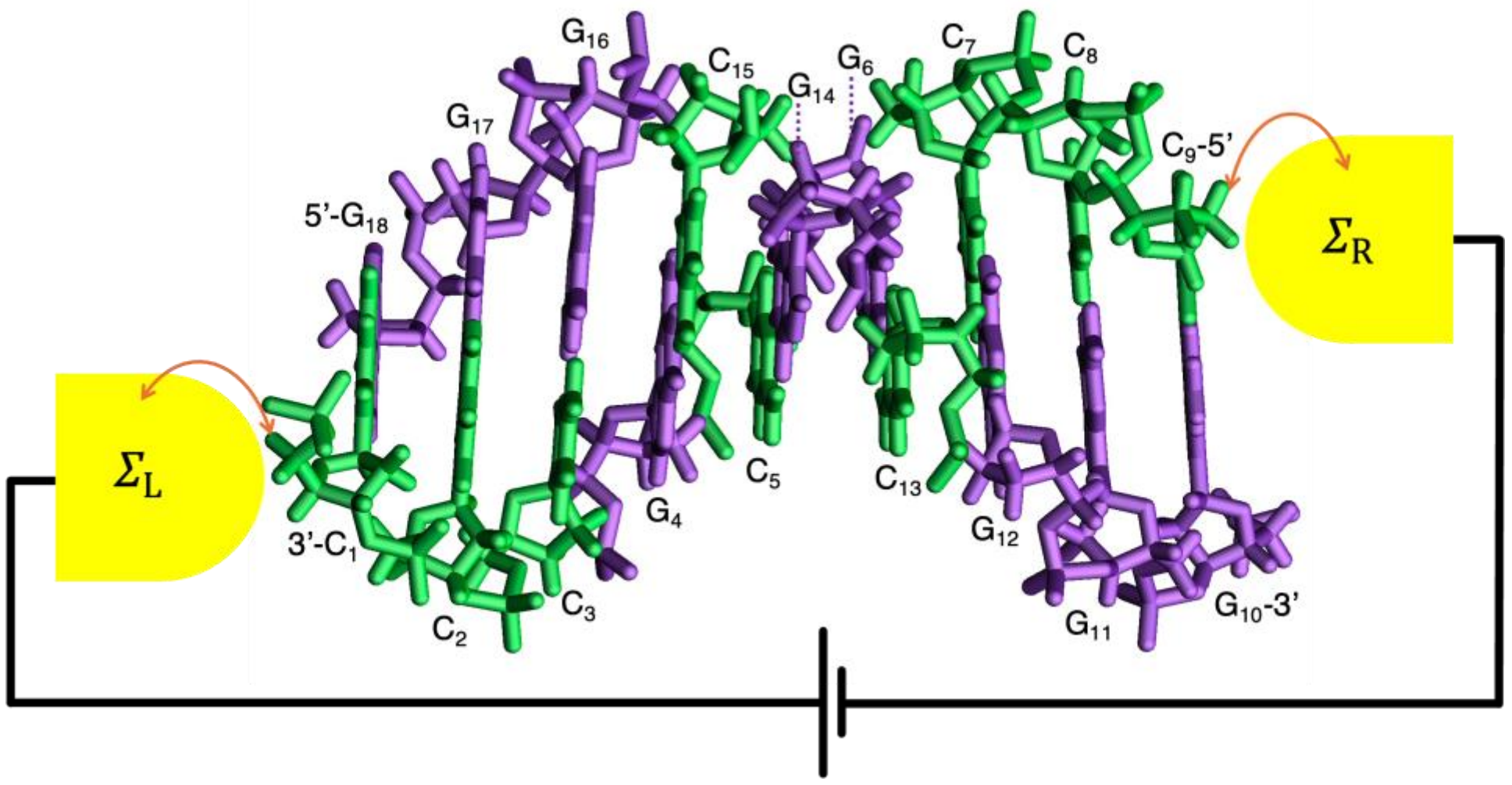

Figure 1 DNA system under study, the guanine nucleotides are colored in purple and the cytosines in green. The contact self-energies are coupled to the terminal cytosines. The figure consist sof 18 partitions, one for each nucleotide as labeled.

The Hamiltonian $H$, as defined in equation (8), is used to compute the retarded Green's function ($G^r$) of the DNA by incorporating the self-energy contributions from both decoherence effects and the contacts,

$$[E - (H + \Sigma_{\text{L}} + \Sigma_{\text{R}} + \Sigma_{\text{D}})]G^r = I \tag{11}$$

Here, $E$ represents the energy, $I$ is the identity matrix, and $H$ is the Hamiltonian discussed in (7)-(10) . The terms $\Sigma_{L(R)}$ refer to self-energies associated with the left (right) contacts, capturing the influence of the contacts attached to the DNA. The final term, $\Sigma_D$, is the decoherence self-energy. In general, the contact self-energy has both a non zero real and imaginary parts. The real part leads to a shift in the energy levels of the DNA but the imaginary part of $\Sigma_{\text{L}}$ and $\Sigma_{\text{R}}$ represent the scattering of electrons between the contacts and the DNA.

For the contact self-energy, we adopted the wide-band limit (WBL) approximation, where the real part of the self-energy is zero and the imaginary part remains constant with energy. Under these assumptions, the only non zero elements of the contact self-energies are given by (see Figure 1),

$$\left[\Sigma_{L(R)}\right]_{(m,k),(m,k)} = -\frac{i\Gamma_{L(R)}}{2}, \qquad m = \text{nucleotides connected to L and R contacts} \tag{12}$$

That is, $\Sigma_{L(R)}$ is a purely complex diagonal matrix. .

The first generation of decoherence probes introduced by Engquist and Anderson [32] and Buttiker [33] use an energy independent decoherence rate given by,

$$\frac{1}{\tau_{m,k}} \propto \frac{\Gamma_{m,k}}{\hbar} \tag{13}$$

where $\tau_{m,k}$ is the scattering time at molecular orbital $k$ of the partitioned nucleotide $m$, $\Gamma_{m,k}$ quantifies the scattering strength, and $\hbar$ is the reduced Planck's constant. Such a model for decoherence yields an unphysically large density of states in the bandgap. Energy dependence can be introduced using an energy dependent model where the decoherence rate [36],

$$\frac{1}{\tau_{m,k}} \propto \frac{\Gamma_{m,k}}{\hbar} e^{-\frac{|E-\tilde{\epsilon}_{k,m}|}{\lambda}} \tag{14}$$

is applied to each partitioned nucleotide. Here, the parameter $\lambda$ determines the decoherence rate decay with energy, $\Gamma_{m,k}$ is the maximum strength of the decoherence rate at $k^{th}$ molecular orbital of the $m^{th}$ partitioned nucleotide. While this decoherence probe model gives the physically meaningful result of small density of states in energy gaps of DNA in the presence of decoherence, it suffers from spurious energy levels that do not exist in the original system. To overcome these drawbacks, we propose to use a decoherence rate that is proportional to the local density of states (LDOS). One possibility is to take,

$$\frac{1}{\tau_{\alpha,i}(E)} \propto LDOS_{(\alpha,i)}(E) \tag{15}$$

where $LDOS_{(\alpha,i)}$ is the localized density of states contributed by orbital $\alpha$ at atom $i$. This model assumes local decoherence, where an electron in orbital $\alpha$ at atom $i$ loses phase coherence without changing its position or orbital when it is reinjected back to the DNA. To permit decoherence that can mix orbitals at an atom, a simple extension of this self-energy is,

$$\frac{1}{\tau_{\alpha,i}(E)} \propto LDOS_i(E) \tag{16}$$

here, $LDOS_i(E) = \sum_\alpha LDOS_{(\alpha,i)}(E)$ is the total density of states at atom $i$, at energy $E$. As a result, an electron in orbital $\alpha$ at atom $i$ can scatter and lose phase coherence to any orbital at the same energy at atom $i$. Alternatively, one can find the basis states corresponding to the molecular orbitals $k$ of the partitioned nucleotide $m$ (the diagonal block $H_{0m}$ of the Hamiltonian). In this basis, the scattering rate,

$$\frac{1}{\tau_{m,k}(E)} = \frac{\Gamma_{m,k}}{\hbar} = \frac{D_o}{\hbar} LDOS_{(m,k)}(E) \tag{17}$$

where $D_o$ is a parameter that determines the strength of coupling between the DNA and the decoherence probe, and hence the decoherence rate. The units of $D_o$ is $eV^2$. Now, the decoherence is limited to a specific molecular orbital $k$ at nucleotide $m$. Like in the case of atomic orbitals, we could let the scattering rate be dependent on the nucleotide and energy but independent of the orbital by taking,

$$\frac{1}{\tau_{m,k}(E)} = \frac{D_o}{\hbar} LDOS_m(E) \tag{18}$$

Here, $LDOS_m(E) = \sum_k LDOS_{(m,k)}(E)$, is the total density of states at nucleotide $m$ at energy $E$. That is, the scattering between the molecular orbitals of each nucleotide is isotropic.

Using the general relationships between scattering rates and self-energy,

$$\frac{1}{\tau_{m,k}(E)} = -\frac{2Im\left[\Sigma_{D(m,k),(m,k)}(E)\right]}{\hbar} \tag{19}$$

where $\Sigma_{D(m,k),(m,k)}$ denotes the $k^{\text{th}}$ diagonal element within the decoherence probe self-energy matrix corresponding to the $m^{th}$ diagonal block. Additionally, the relationship between LDOS at nucleotide $m$ and the Green's function is,

$$LDOS_m(E) = \sum_k LDOS_{(m,k)}(E) = -\frac{1}{\pi}\sum_k Im\left[G^r_{(m,k),(m,k)}(E)\right] \tag{20}$$

Combining equations (18)-(20), we can define the decoherence self-energy as,

$$Im[\Sigma_{D_{(m,k),(m,k)}}(E)] = \frac{D_o}{2\pi}\sum_k Im\left[G^r_{(m,k),(m,k)}(E)\right] \tag{21}$$

The real part of the self-energy to related to the imaginary part via the Kramers-Kronig relationship:

$$Re\left[\Sigma_{D_{(m,k),(m,k)}}(E)\right] = \frac{1}{\pi}P\int \frac{Im\left[\Sigma_{D_{(m,k),(m,k)}}(E')\right]}{E'-E}dE' \tag{22}$$

where $P$ is the Cauchy principal value. Using equation (21) in equation (22), we have,

$$Re\left[\Sigma_{D_{(m,k),(m,k)}}(E)\right] = \frac{D_o}{2\pi}\sum_k \frac{1}{\pi}\,P\int \frac{Im\left[G^r_{(m,k),(m,k)}(E')\right]}{E'-E}dE' \tag{23}$$

Using the Kramers-Kronig relationship between the real and imaginary parts of $G^r$,

$$Re\left[\Sigma_{D_{(m,k),(m,k)}}(E)\right] = \frac{D_o}{2\pi}\sum_k Re\left[G^r_{(m,k),(m,k)}(E)\right] \tag{24}$$

Now, by applying equations (21) and (24) to include both the real and imaginary parts of $\Sigma_{D_{(m,k),(m,k)}}$, the non-zero elements of the self-energy due to the decoherence probe are:

$$\Sigma_{D_{(m,k),(m,k)}}(E) = \frac{D_o}{2\pi}\sum_k G^r_{(m,k),(m,k)}(E) \tag{25}$$

That is, the decoherence self-energy at molecular orbital $k$ on nucleotide $m$ is an equally weighted sum of contributions from all other molecular orbitals at the same nucleotide at energy $E$. The Green's function equation for the system in equation (11) has now been fully defined. This is shown pictorially in Figure 2, where every nucleotide $m$ is connected to a separate decoherence probe at every orbital.

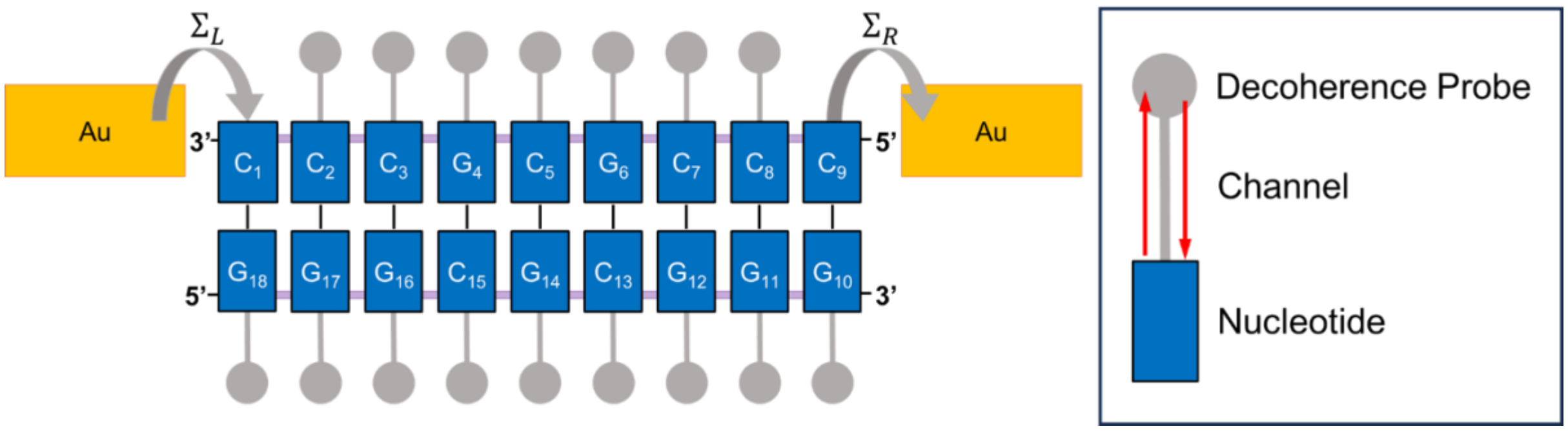

Figure 2 Schematic representation of the decoherence probe model applied to the dsDNA. Each block is attached to a decoherence probe. The blocks are defined as nucleotides (base + backbone). The purple lines are to emphasize that the DNA is a connected chain through the backbone.

We also define the total Density of States (DOS) over the entire device as the sum of the LDOS at all nucleotides at energy $E$,

$$DOS(E) = \sum_m LDOS_m(E) = -\sum_m \sum_k \frac{Im\left[G^r_{(m,k),(m,k)}(E)\right]}{\pi} \tag{26}$$

The scattering rate $\Gamma_{m,k}(E)/\hbar$ between the DNA and the $m^{th}$ decoherence probe is defined by,

$$\Gamma_m(E) = \Gamma_{m,k}(E) \equiv -2Im\left(\Sigma_{D(m,k),(m,k)}(E)\right) = \frac{\hbar}{\tau_{m,k}} = \frac{\hbar}{\tau_m} \tag{27}$$

Note that both $\Gamma_{m,k}$ and the scattering rate $\hbar/\tau_{m,k}$ do not depend on the orbital index $k$ because the quantity $\Sigma_{D(m,k),(m,k)}$ in equation (25) has been summed over the orbital index $k$ and is hence independent of the orbital at nucleotide location $m$. So, in Section 4, we will simply refer to them by $\Gamma_\mathrm{m}$ and $\tau_\mathrm{m}$. In the low-bias regime, the current per unit energy at the $m^{th}$ probe, which includes the decoherence connected to the $m^{th}$ nucleotide, left contact and right contact) is given by:

$$J_m(E) = \frac{2q}{h}\sum_{n=1}^{N} T_{mn}(E)[f_m(E) - f_n(E)] \tag{28}$$

where $q$ is the elementary charge, $h$ is Planck's constant. Note that decoherence probes are not connected to the probes where we have the left and right contact though it can be trivially generalized to have decoherence probes at all partitioned nucleotides. $T_{mn}$ is the transmission between probes $m$ and $n$, and it is defined via the well-known formula,

$$T_{mn}(E) = \mathrm{Tr}\left[\Gamma_m(E) \times G^r_{m,n}(\mathrm{E}) \times \Gamma_n(E) \times G^a_{n,m}\right] \quad (29)$$

Here $G^r_{m,n}$ is the Green's function submatrix with row indices corresponding to all orbitals in nucleotide $m$ and column indices corresponding to all orbitals in nucleotide $n$. $\Gamma_m = -2Im\left(\Sigma_{D(m,k),(m,k)}(E)\right)$ is a square matrix with dimension equal to the number of molecular orbitals in nucleotide $m$. The advanced Green's function is given by $G^a = (G^r)^\dagger$ , and $f_m(E) = \left[1 + \exp\left(\frac{E-\mu_m}{k_B T}\right)\right]^{-1}$ at probe $m$, $k_B$ is the Boltzmann constant, and $T$ is the temperature. In the left and right contacts, $f_m(E)$ is the Fermi function. However, in the decoherence probes, while $f_m$ looks like the Fermi distribution function, it is really a different function because $\mu_m$ is a function of energy. While we do not know the parameter $\mu_m$ a priori, we know the chemical potential of the left ($\mu_L$) and right ($\mu_R$) contacts. When the applied bias between the left and right contacts is small and the system is in linear response regime, $f_m(\mathrm{E}) \approx f_{eq}(E) + (\mu_m - \mu)\left(-\frac{\partial f_{eq}(E)}{\partial E}\right)$, where $\mu$ is the equilibrium chemical potential in the left (L) and right (R) contacts at zero bias and $f_{eq}(E) = \left[1 + \exp\left(\frac{E-\mu}{k_B T}\right)\right]^{-1}$. Then, the equation for $J_m(E)$ simplifies to:

$$J_m(\mathrm{E}) = \frac{2q}{h}\sum_{n=1}^{N} \tilde{T}_{mn}(E)\,[\mu_m - \mu_n], \quad m: 1 \to N \quad (30)$$

where,

$$\tilde{T}_{mn}(E) = T_{mn}(E)\left(-\frac{\partial f(E)}{\partial E}\right)_{eq} \quad (31)$$

As the decoherence probe $m$ mimics decoherence at nucleotide $m$ without carrying a net current in the fictitious decoherence probe $m$, $J_m(E) = 0$. An elastic model for scattering between the DNA and decoherence probe $m$ is assumed. As a result, the sum of currents in the left and right contacts at each energy is zero, $J_L(\mathrm{E}) + J_R(\mathrm{E}) = 0$, which gives [41],

$$J_L(\mathrm{E}) = \frac{2q}{h} T_{eff}(E) \left(-\frac{\partial f(E)}{\partial E}\right)_{eq} [\mu_L - \mu_R] \tag{32}$$

where,

$$T_{eff} = T_{LR} + \sum_{m,n}^{N_B} T_{Lm} W_{mn}^{-1} T_{nR} \tag{33}$$

$W_{mn}^{-1}$ is the inverse of $W_{mn} = (1 - R_{mm})\delta_{mn} - T_{mn}(E)(1 - \delta_{mn})$, $R_{mm}$ is the reflection probability at probe $m$, which is given by $R_{mm} = 1 - \sum_{m \neq n}^{N} T_{mn}$, and $N_B$ is the number of decoherence probes. In equation (33), the first term represents the direct transmission from the left to the right contact. The second term captures the contribution arising and electron traveling from the left contact scattering into and out of the decoherence probes before traveling to the right contact. Now the total current flowing at the left contact $I_L$ is the integral of $J_L(\mathrm{E})$,

$$I_L = \int_{-\infty}^{\infty} J_L(\mathrm{E})\, dE = \frac{2q}{h} (\mu_L - \mu_R) \int_{-\infty}^{\infty} T_{eff}(E) \left(-\frac{\partial f(E)}{\partial E}\right)_{eq} dE \tag{34}$$

Then, the linear response conductance when the equilibrium chemical potential is $\mu$ is,

$$G(\mu) = \frac{I_L}{(\mu_L - \mu_R)/q} = \frac{2q^2}{h} \int_{-\infty}^{\infty} T_{eff}(E) \left(-\frac{\partial f(E)}{\partial E}\right)_{eq} dE \tag{35}$$

At zero temperature, the above expression becomes,

$$G(\mu) = \frac{2q^2}{h} T_{eff}(\mu) \tag{36}$$

To find the transmission, the calculation proceeds following an independent iterative process at each energy. In Iteration 1, equation (11) is solved using only the contact self-energies $\Sigma_{\mathrm{L}}$ and $\Sigma_{\mathrm{R}}$ in equation (12), with $\Sigma_{\mathrm{D}}$ set to zero, which is the coherent limit. To describe the iterative process for iteration 2 onwards, it is simpler if we rewrite equation (11) as

$$\left[E - \left(H + \Sigma_{\mathrm{L}} + \Sigma_{\mathrm{R}} + \Sigma_{D(j+1)}\right)\right] G_{(\mathrm{j}+1)}^{r} = I \quad (j \geq 1)$$

where the Green's function in iteration $j+1$ ($G^r_{(j+1)}$) depends on the Green's functions and LDOS found in previous iterations via the following equations which uses linear mixing,

$$Im\left[\Sigma_{D_{(m,k),(m,k)}}(E)\right]_{(j+1)} = \alpha \cdot \frac{D_o}{2\pi} \sum_k Im\left[G^r_{(m,k),(m,k)}(E)\right]_{(j)} + (1-\alpha) \cdot \frac{D_o}{2\pi} \sum_k Im\left[G^r_{(m,k),(m,k)}(E)\right]_{(j-1)}$$
$$Re\left[\Sigma_{D_{(m,k),(m,k)}}(E)\right]_{(j+1)} = \alpha \cdot \frac{D_o}{2\pi} \sum_k Re\left[G^r_{(m,k),(m,k)}(E)\right]_{(j)} + (1-\alpha) \cdot \frac{D_o}{2\pi} \sum_k Re\left[G^r_{(m,k),(m,k)}(E)\right]_{(j-1)} \quad (37)$$

For $j = 1$ alone, the Green's function on the right-hand side at iteration $j-1=0$ is set equal to zero. Afterwards, equations (26), (27), (29) and (33) are used to evaluate the total DOS, $\Gamma_m(E)$, and the effective transmission $T_{eff}(E)$, respectively. The total DOS and $T_{eff}$ are compared to their previous iteration using the criteria

$$\frac{\left|DOS_{(j)} - DOS_{(j-1)}\right|}{DOS_{(j-1)}} \times 100 < \eta_{DOS},$$
$$\frac{\left|T_{eff\,(j)} - T_{eff\,(j-1)}\right|}{T_{eff\,(j-1)}} \times 100 < \eta_{Tr} \quad (38)$$

This iterative process is repeated until convergence is reached as shown in the flowchart in Figure 3. The convergence criteria per energy point requires both the total DOS and Transmission differences in equation (38) to be below their threshold values $\eta_{DOS}$ and $\eta_{Tr}$, respectively. Throughout the manuscript, we set $\eta_{DOS} = \eta_{Tr} = 0.1\%$, using a more stringent threshold of 0.001% produced negligible changes in transmission (see SI, Section 8).

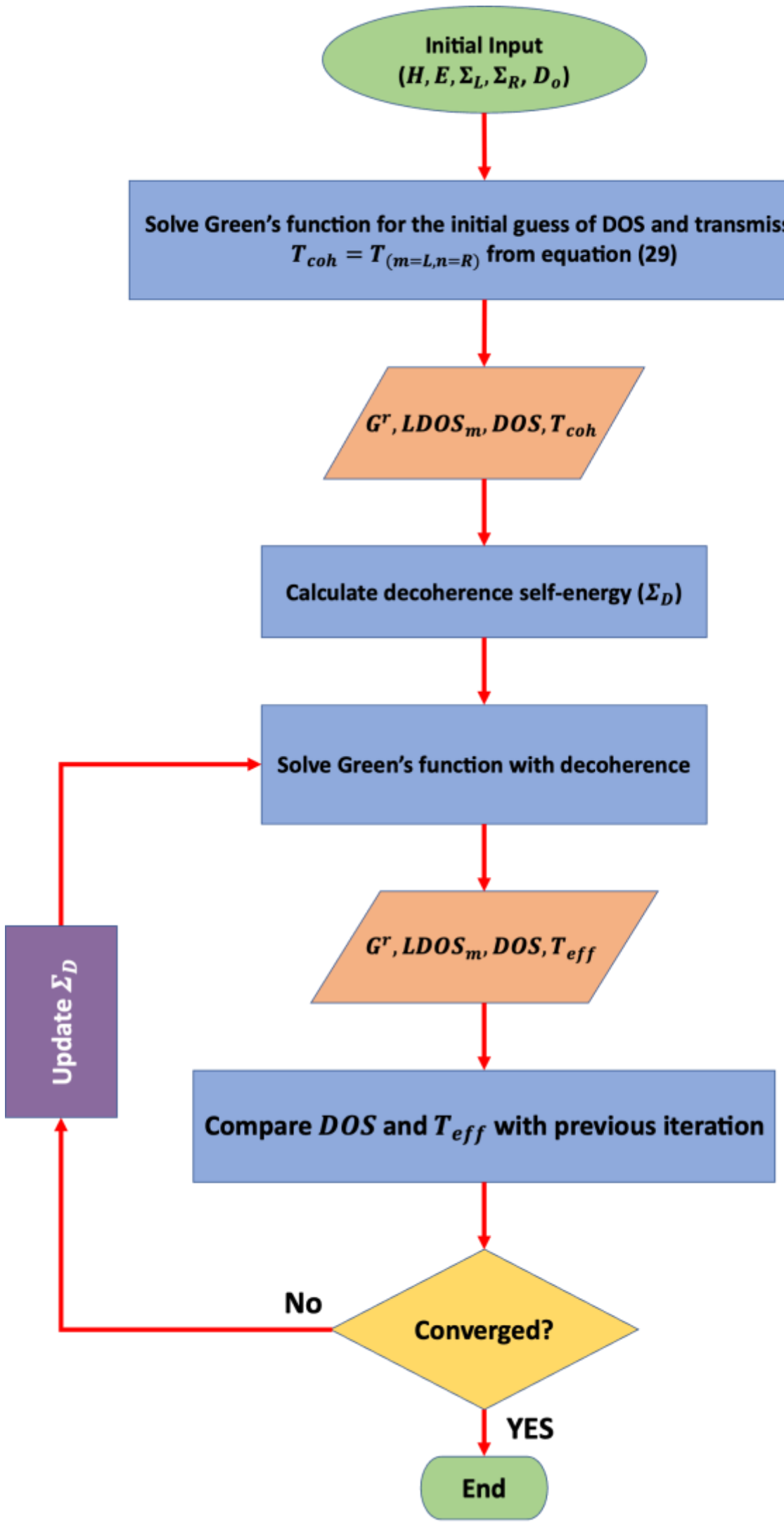

Figure 3 Flowchart of the DOS-weighted decoherence models.

## 2.2 Relationship To Other Decoherence Probe Models

We summarize the differences between the different decoherence model implementations in Figure 4. In the previous implementations, the decoherence was first defined as Energy-Independent (*E-indep*), with $\Gamma_m(\mathrm{E}) = \Gamma_\mathrm{B}$ (constant) independent of energy at nucleotide $m$. The drawback of using an *E-indep* model is that it leads to excessive broadening of the DOS across all energies. This yields unphysically large transmission in energy gaps, as all energy levels are uniformly broadened regardless of the transmission energy value., which is unphysical. To rectify the excessive broadening of DOS, the energy-dependent (*E-dep*) decoherence model was  proposed in reference [36], with $\Gamma_{m,k}(E) = \Gamma_B \exp\left(-\frac{|E-\tilde{\varepsilon}_{k,m}|}{\lambda}\right)$, which is the same as equation (14) with $\Gamma_{m,k} = \Gamma_B$. The decoherence rate $\Gamma_{m,k}/\hbar$ , reaches its peak at the $k^{th}$ molecular orbital energy of partioned nucleotide $m$, $\tilde{\epsilon}_{k,m}$ (obtained from the diagonal Hamiltonian block $H_{0m}$), and decays exponentially with decay constant $\lambda$. While this decoherence model resulted in greatly reduced DOS in energy gaps of the DNA, this approach suffered from spurious peaks appearing in energy gaps because the molecular orbital energy $\tilde{\epsilon}_{k,m}$ was calculated by abruptly cutting the bonds in the backbone of nucleotide $m$ and not accounting for interaction with other bases (see Section 5 and reference [36] for a more detailed discussion). Further, the *E-dep* model adds an additional parameter $\lambda$ that needs to be tested and fitted with experiments. The proposed model in this work addresses these previous issues by directly relating the strength of the decoherence probes to the DOS of the full DNA system with contacts. Compared to the *E-dep* model that relied on two parameters $\Gamma_B$ and $\lambda$, it reduces the parameters to a single variable $D_o$, which reflects the strength of decoherence. Further, the decoherence is now dependent on the DOS of the entire DNA plus contact system rather than the energies of the artificially partitioned molecular orbitals $\tilde{\epsilon}_{k,m}$, thus, it solves the spurious peaks in the HOMO-LUMO and other energy gaps of the DNA. This new model, however, requires more computational time since it goes through an iterative process to reach convergence.

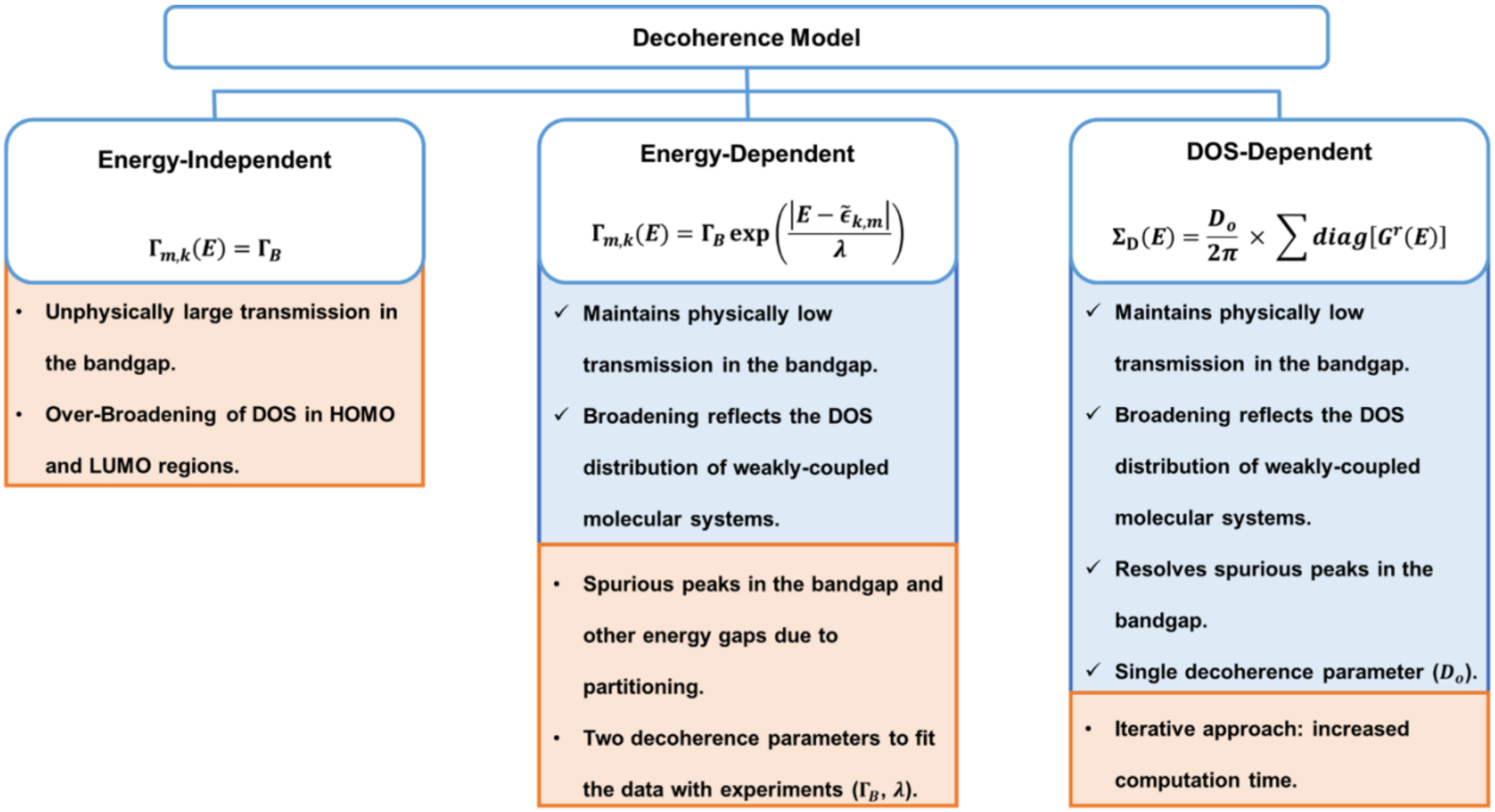

Figure 4 Comparison of the different decoherence models.

# 3 Effect of Including the Real Part of the Decoherence Probe Self-Energy

Our previous *E-dep* model showed that the inclusion of the real and the imaginary parts of the self-energy yields only small differences from including only the imaginary part, in calculating the transmission near the HOMO. However, its inclusion was crucial to yield the correct integral of the DOS [36,37]. In this section, we look at the effect of including the real and imaginary parts of the DOS-weighted decoherence probe self-energy on the transmission. We start with a single level system $\epsilon = 0$ eV, and plot the real and imaginary parts of $\Sigma_D$ for $D_o = 0.1\ eV^2$ in Figure 5(a). The real part shifts the onsite potential while the imaginary part corresponds to the inverse of the scattering rate into the decoherence probe broadens the energy level. We apply the decoherence self-energy to the single level system and the resulting DOS is plotted in Figure 5 (b). If the real part of the self-energy is neglected ($\Sigma_D(E) = iIm[\Sigma_D(E)]$), the self-energy becomes defined by only the imaginary part of $\mathrm{G^r}$. In this case, the blue dotted curve in Figure 5(b) shows that the DOS becomes narrower compared to the result with the complex self-energy (black curve in Figure 5(b)). The difference increases as the decoherence strength ($D_o$) grows (see SI Figure S1), and the effect becomes more pronounced when we extend the system to include additional energy levels, as discussed below.

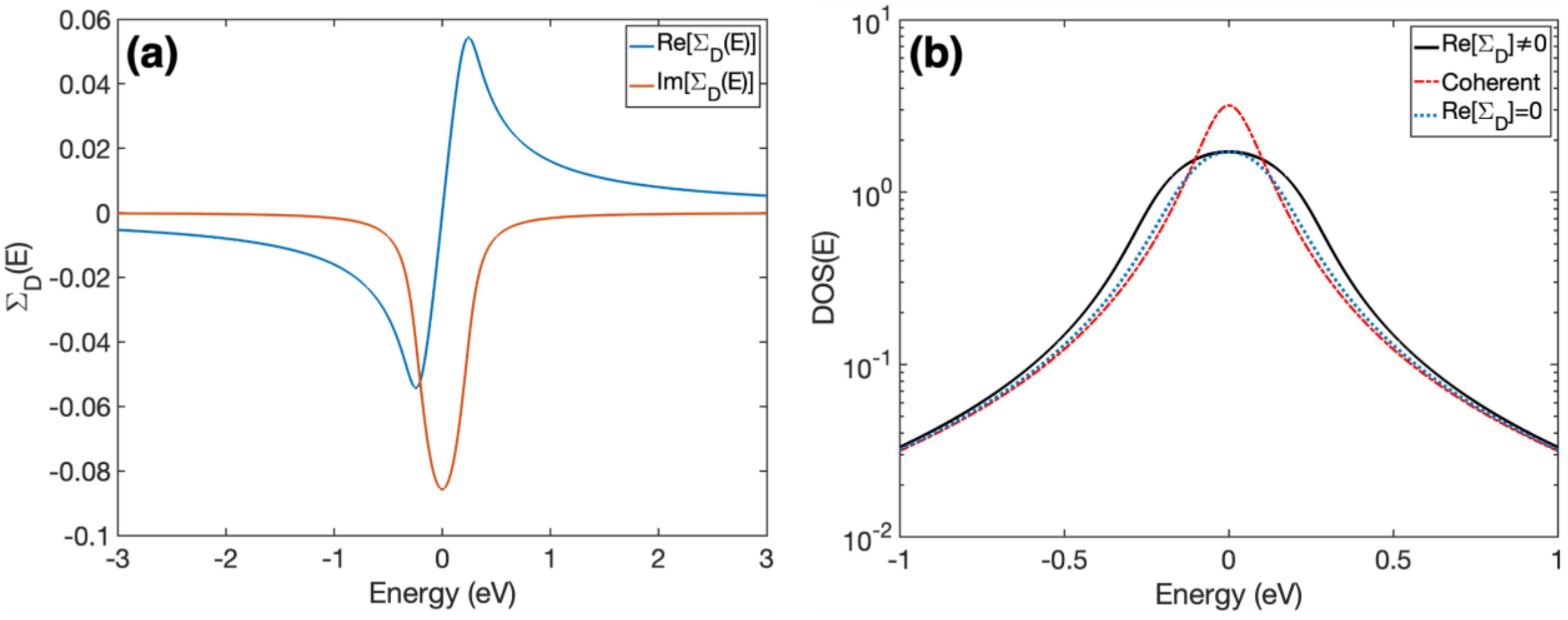

Figure 5 Single level system results. (a) The real and imaginary parts of the decoherence self-energy. (b) Density of states for the full complex decoherence self-energy and only the imaginary part of the self-energy, the coherent result is also shown ($D_o = 0\ eV^2$). The calculation parameters are $D_o = 0.1\ eV^2, \Gamma_{L(R)} = 100\ meV$. The integral of only the coherent and the full $Full\ \Sigma_D$ curves is one.

Next, we extended the system to simulate three nucleotides in a DNA strand. We generated a model Hamiltonian that includes only the HOMO and LUMO of each nucleotide, following our approach in previous work [36] and discussed in Section 2 of the SI. For the case with the full self-energy (a) ($Full\ [\Sigma_D(E)]$), the $\int DOS\ dE = 6$, which is the number of states in the system (without the spin multiplicity). However, when we exclude the real part ($Re[\Sigma_D(E)] = 0$), the integral of the DOS yields 5.45, which is incorrect. Further, Figure 6 shows that including the $Re[\Sigma_D(E)]$ broadens the DOS and transmission peaks over a wider energy range.

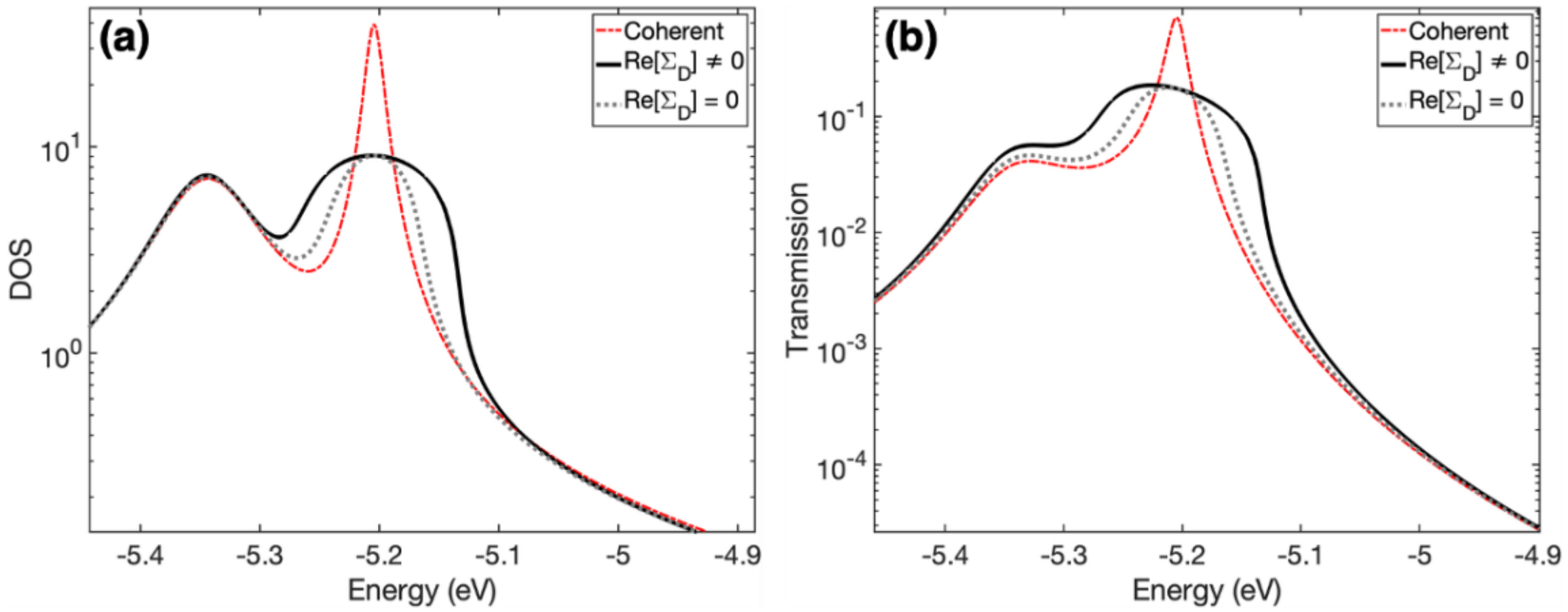

Figure 6 DOS and transmission of the single-strand model Hamiltonian 5'-CCC-3' comparing the use of the decoherence self-energy with and without the $Re[\Sigma_D(E)]$ with $D_o = 0.01\ \mathrm{eV}^2$, the coherent transmission is also included as reference. $\Gamma_{L(R)} = 100\ \mathrm{meV}$,

We also compared the results obtained by calculating the decoherence self-energy by two numerical methods: (1) calculation of decoherence self-energy by directly substituting for $\mathrm{G}^\mathrm{r}$ obtained by solving equation (11) in the RHS of equation (25), versus (2) using the Kramers-Kronig relation (equation (22)) to calculate the real-part of the decoherence self-energy separately, by substituting for the imaginary part of the decoherence self-energy on the RHS of equation (21). Both approaches yield the same result (See Figure S3 in SI). While both numerical methods require self-consistent iterative calculations to reach convergence (Figure 3), method 1 took more than 16 times less computational time than method 2. This is because the Kramers-Kronig relation requires evaluating $Im\left[\Sigma^r_{D(m,k),(m,k)}(E')\right]$ over a wide energy range in variable $E'$ and performing the integral repeatedly at every energy $E$ in equation (22). Based on the results and analysis of this section, in the rest of the manuscript, we apply method 1 as described by equation (25) on a full DNA plus contacts system.

# 4 Results and Comparison

We modeled a GC-rich 9 base pair ds-DNA 3'-CCCGCGCCC-5' adopted from [36]. We used the following parameters for the DOS-weighted model: $D_o = [0.001, 0.01, 0.025, 0.1]\ \mathrm{eV}^2$, and convergence criteria $\eta_{DOS} = \eta_{Tr} = 0.1\%$. For the *E-dep* decoherence model, we used $\Gamma_B = 100\ meV$, and $\lambda = 50\ meV$. For all models, we employed nucleotide partitioning and set $\Gamma_L = \Gamma_R = 600\ meV$ at the 3'- and 5'- end cytosines of this system ($C_1$ and $C_9$ in Figure 1), to be consistent with the previous study [36].

We selected a range of $D_o$ values to examine its influence on the transmission and DOS spectra. In reality, decoherence rates depend on several factors, including temperature, strand length, solvent environment, and vibrational modes. The $D_o$ parameter in our model is intended to capture the combined effect of these contributions to dephasing within the elastic scattering limit. A precise determination of scattering rates requires coordinated theoretical and experimental work. In the present study, we instead explore a broad range of $D_o$ values and compare the resulting trends with available experimental data to estimate a reasonable range for $D_o$.

The calculations converge and the transmission plot in Figure 7 shows the different models, including the energy-independent decoherence transport results. We observe that the DOS-weighted decoherence model behavior with respect to $D_o$ has similarities to both the *E-indep* and *E-dep* models. For instance, at low $D_o$ of $0.001\ \text{eV}^2$, the transmission becomes closer to the coherent limit, something that the *E-dep* model mimics for low $\Gamma_B$ and $\lambda$ [36]. At moderate $D_o = 0.01\ \text{eV}^2$ and $0.025\ \text{eV}^2$, increasing $D_o$ increases the broadening of the DOS and hence increases the off-resonant transmission. One key difference to the *E-dep* model appears at the large $D_o = 0.1\ \text{eV}^2$, where the transmission peaks display significant merging due to the larger decoherence broadening. The transmission profile at high DOS regions that are energetically close (-5.1 eV < E < -4.8 eV and E <-5.6 eV) starts showing resemblance to the *E-indep* model (black and green curves in Figure 7). Additionally, the $Re[\Sigma_D]$ shifts the onsite potentials, causing the high transmission values to extend over a broader energy range (black curve at -5.6 eV < E < -5.4 eV and -4.7 eV < E < -4.6 eV), with the transmission decay into the HOMO-LUMO gap now occurring 138 meV above the HOMO.

We attribute this difference between *DOS-weighted* and *E-dep* decoherence models to the distinct ways decoherence is introduced at the nucleotide: the *E-dep* model acts on the localized orbitals of each partitioned nucleotide $m$ and decays with energy, whereas the *DOS-weighted* model relies on the LDOS at each nucleotide calculated for the full system with contacts. A detailed analysis of how this affects the energy-dependent broadening and its relation to the system eigenstates and transmission is provided in Section 5. Further, the *DOS-weighted* decoherence model maintains the transmission decay into the HOMO-LUMO gap (E > -4.7 eV), an aspect that the *E-indep* decoherence model misses. The qualitative behavior of transmission versus energy in the *E-dep* model at other energy gaps (e.g., E~-5.5 eV) is also preserved in the *DOS-weighted* model. Additionally, we find that the DOS-weighted model reproduces the experimental low-bias conductance range by varying a single decoherence parameter ($D_o$). Further details of the conductance calculations are provided in Section 4 of the SI. Moreover, DNA sequence influences the distribution of energy levels along the strand. To demonstrate that the conclusions above hold for a broader set of sequences, we applied the *DOS-weighted* decoherence model to a sequence containing six AT base pairs neighboring three central GC pairs. Additional details are provided in Section 5 of the SI.

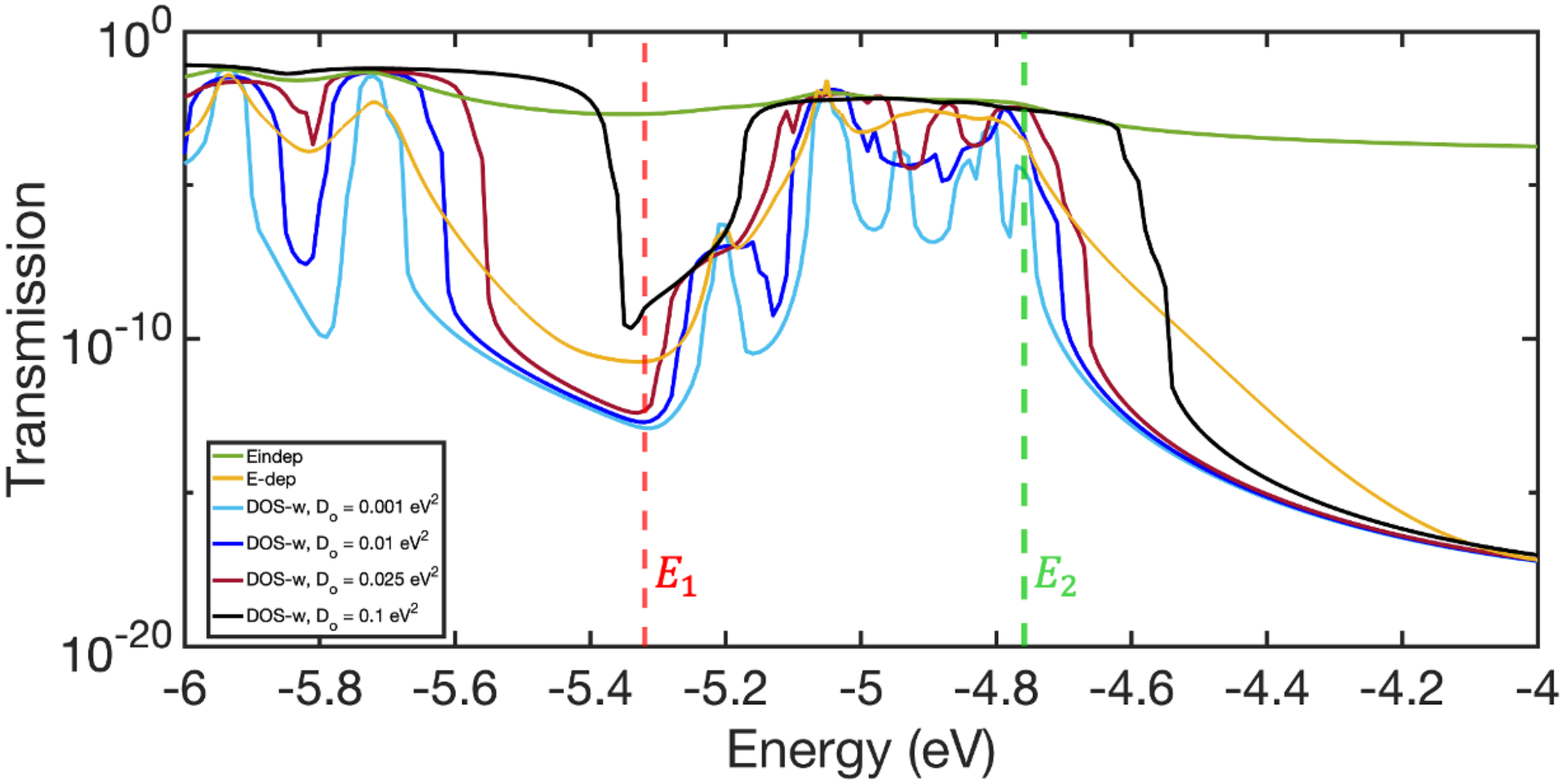

Figure 7 Transmission of the DNA using different decoherence models in the HOMO region going to the HOMO-LUMO gap. The DOS-weighted decoherence probe model is abbreviated as DOS-w in the legend. The vertical dashed lines correspond to $E_1$= -5.32 eV and $E_2$ = HOMO = -4.759 eV, which are further discussed in the text.

## 4.1 Decoherence Rates and Coherence Lifetime

The converged decoherence scattering rate in this model depends on the value of $D_o$, the sequence, and energy. We find that at small values of $D_o$, the scattering rate has a peak at the energies close to the location where the coherent DOS is large. However, as the value of $D_o$ increases, substantial shifts induced by the real part of the self-energy, together with contributions from neighboring shifted states, modify the local DOS, causing both the spatial position of the DOS peak and the associated scattering rate peak at a fixed energy to differ from those in the lower-$D_o$ regime. We will illustrate below that the DOS peak may occur at a nucleotide and at an energy distinct from those at lower values of $D_o$ (see Figure 8). The DOS has a peak close to the HOMO energy of $E_2 = -4.759\ eV$ (see green line in Figure 7 and Figure 8 (a)) at nucleotide $G_{16}$ in the coherent limit. At $D_o = 0.001$ and $0.01$ eV$^2$, the DOS and correspondingly the decoherence rate continues to have a peak very close to energy $E_2 = -4.759\ eV$ (see Figure 8 (b-c)). However, as the value of $D_o$ increases to $0.1$ eV$^2$, the peak in DOS close to $E \sim E_2 = -4.759\ eV$ shifts a few base pairs away to $G_{14}$ (see Figure 8 (d)). The DOS and decoherence scattering rate at base $G_{16}$ have peaks at energies close to -4.758 eV, -4.747 eV for $D_o = 0.001$ and $0.01$ eV$^2$. However, as the value of $D_o$ increases to $0.1$ eV$^2$, the shape of the DOS changes significantly at the same base ($G_{16}$) and the peaks of the DOS and scattering rate are at -4.67 eV, which is 111 meV away from the HOMO of the coherent limit. G14 now has the DOS peak closest to HOMO at -4.75 eV. This shift occurred because of the increased level shift and broadening of the state at $E \approx -4.85$ eV, which lies energetically near the HOMO and is spatially localized at $G_{14}$, which is shown in Figure 8 (b-d).

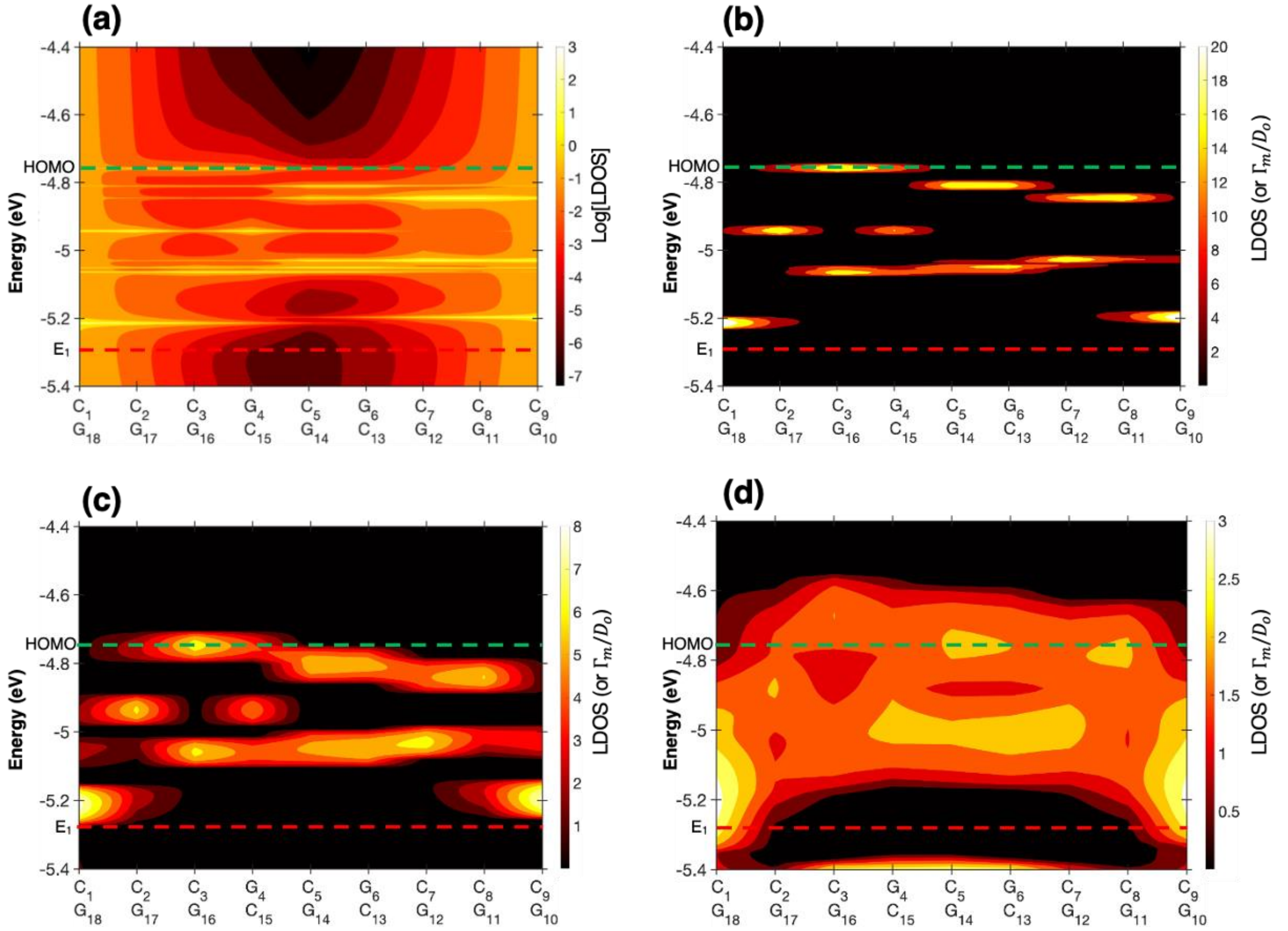

Figure 8 2D DOS plots for (a) coherent transport, (b) $D_o = 0.001\ eV^2$,(c) $D_o = 0.01\ eV^2$, and (d) $D_o = 0.1\ eV^2$. Note that for clarity, $log_{10}(DOS)$ was used to plot the coherent case. Note that the 2D DOS plot also represents $\Gamma_m/D_o$.

We illustrate the above points further by considering two energy values: $E_1$= -5.32 eV where the transmission and DOS are low in the coherent limit (see red line in Figure 7 and Figure 8) and $E_2$ = HOMO = -4.759 eV. Figure 9 shows that $\Gamma_m$ (and hence $\tau_m$) can vary by orders of magnitude along the strand, at both energies. Further, as the DOS at $E_2$ = HOMO is significantly higher than at $E_1$ which is off resonance, the decoherence values across the DNA strand are on average two orders of magnitude larger at $E_2$. Qualitatively, we notice that $\Gamma_m$ increases by approximately ten times when the value of $D_o$ increases ten times from the lower value of 0.001 $\mathrm{eV}^2$ to the moderate value of 0.01 $\mathrm{eV}^2$. However, when $D_o$ increases further by ten times to 0.1 $eV^2$, we find that $\Gamma_m$ can increase by more than ten times. At the highest considered value of $D_o = 0.1$ eV$^2$, the spatial distribution of the scattering rate changes. *Such an effect is seen because of the energy shift induced by decoherence scattering, at which the density of states peaks (the energy shift occurs due to the real part of self-energy) and because of decoherence scattering induced broadening.* For instance, at $E \sim E_1$, the $\Gamma_m$ value at nucleotide $G_{10}$ increases from 0.13 meV (for $D_o = 0.01\ eV^2$) to 88 meV (for $D_o = 0.1\ eV^2$), overcoming $C_8$ as the nucleotide with the peak scattering rate. Similarly, the scattering rate at $G_{11}$ overcomes the value at its complementary base, $C_8$, at $D_o = 0.1\ eV^2$.

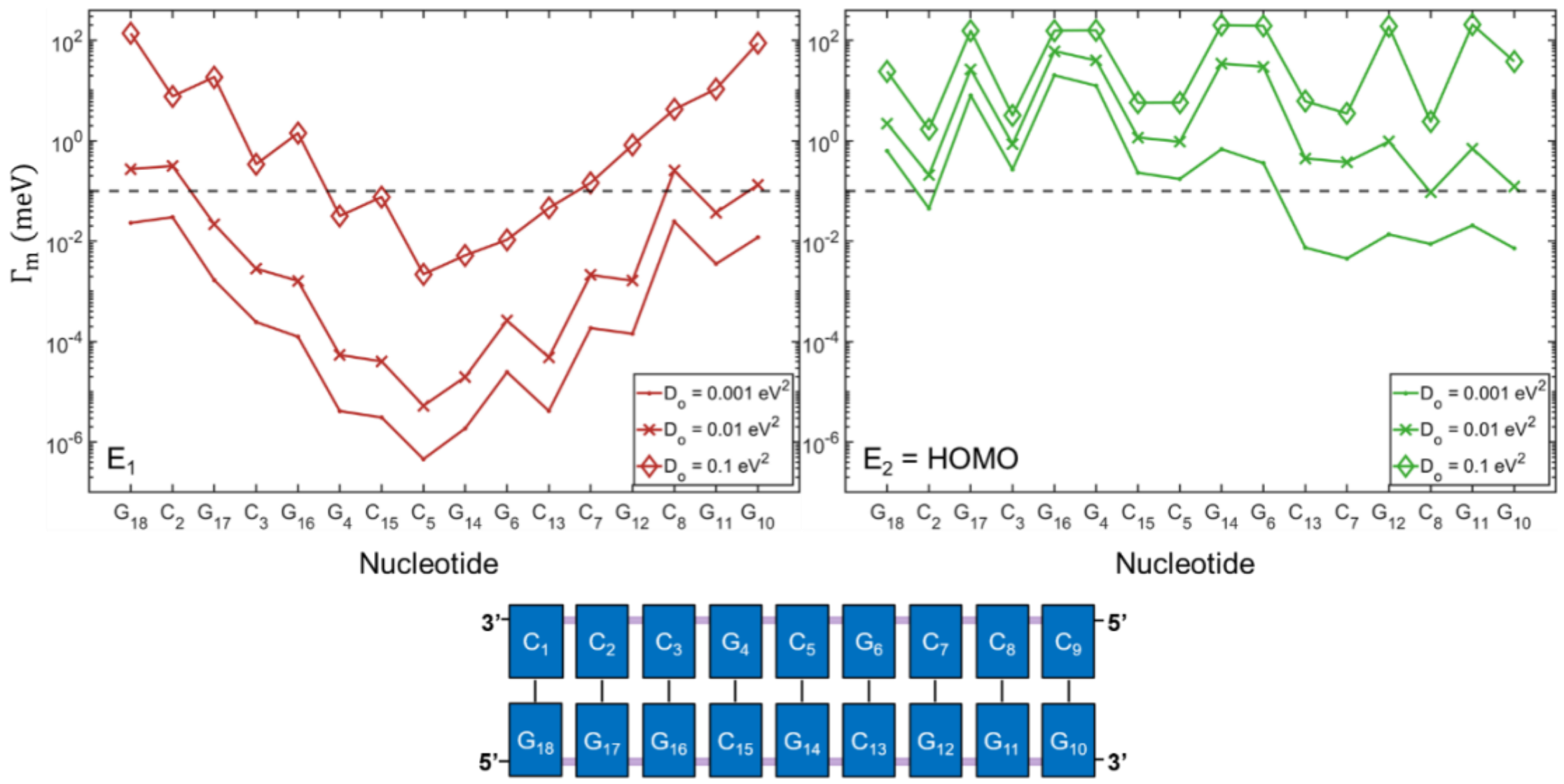

Figure 9 Converged decoherence rates for $E_1 = -5.32\ eV$ (left) and $E_2 = -4.759$ (right) at different $D_o$ values.The contact sites $C_1$ and $C_9$ are omitted from the x-axis because they do not have decoherence probes. Bottom inset shows the ds-DNA system under study. The plots contain the nucleotide resolved data from Figure 8 at energies $E_1$ and $E_2$.

We also show the overall distribution of $\Gamma_m$ values for each $D_o$ in Figure S8 of the SI and list the range of values obtained in Table 1 and Table 2. At $E_1$, we can see that the qualitative observations made above are quantitatively reflected in the values of $\Gamma_m$. For $D_o = 0.001\ eV^2$, the lowest decoherence strength considered, the scattering rate is small at all nucleotides except one, and ranges from $4.6 \times 10^{-7}$ meV to a maximum of 0.87 meV. At $D_o = 0.01\ eV^2$, the values of the scattering rate are 10 times larger. As for $D_o = 0.1\ eV^2$, most of the nucleotides have considerably large values of $\Gamma_m$ (Figure S8), with the median increasing by more than 300 compared to $D_o = 0.01\ eV^2$ (see Table 1).

Table 1 The minimum, maximum and median decoherence values $\Gamma_m$ and corresponding coherence lifetime $\hbar/\Gamma_m$ at $E_1 = -5.32\ eV$.

| $\boldsymbol{D_o}$ (eV$^2$) | $\boldsymbol{\Gamma_{min}}$ (meV) | $\boldsymbol{\Gamma_{max}}$ (meV) | $\boldsymbol{\Gamma_{med}}$ (meV) |
|---|---|---|---|
| 0.001 | $4.6\times10^{-7}$ ($\tau_{max}$: 1.4 µs) *Loc: $C_5$* | 0.03 ($\tau_{min}$: 21.9 ps) *Loc: $C_2$* | $1.6\times10^{-4}$ ($\tau_{med}$: 4 ns) |
| 0.01 | $5.3\times10^{-6}$ ($\tau_{max}$: 124 ns) *Loc: $C_5$* | 0.31 ($\tau_{min}$: 2.1 ps) *Loc: $C_2$* | $1.88\times10^{-3}$ ($\tau_{med}$: 356 ps) |
| 0.025 | $1.8\times10^{-5}$ ($\tau_{max}$: 37 ns) *Loc: $C_5$* | 1.01 ($\tau_{min}$: 650 fs) *Loc: $G_{18}$* | 0.007 ($\tau_{med}$: 95.2 ps) |
| 0.05 | $2\times10^{-4}$ ($\tau_{max}$: 3 ns) *Loc: $C_5$* | 48.5 ($\tau_{min}$: 13.6 fs) *Loc: $G_{18}$* | 0.03 ($\tau_{med}$: 24.4 ps) |
| 0.1 | $2.2\times10^{-3}$ ($\tau_{max}$: 299 ps) *Loc: $C_4$* | 137.4 ($\tau_{min}$: 4.8 fs ) *Loc: $G_{18}$* | 0.6 ($\tau_{med}$: 1.37 ps) |

Table 2 The minimum, maximum and median decoherence values $\Gamma_m$ and corresponding coherence lifetime $\hbar/\Gamma_m$ at $E_2 = -4.759$ eV (HOMO).

| $\boldsymbol{D_o}$ (eV$^2$) | $\boldsymbol{\Gamma_{min}}$ (meV) | $\boldsymbol{\Gamma_{max}}$ (meV) | $\boldsymbol{\Gamma_{med}}$ (meV) |
|---|---|---|---|
| 0.001 | 0.005 ($\tau_{max}$: 146 ps) *Loc: $C_7$* | 20 ($\tau_{min}$: 33 fs) *Loc: $G_{16}$* | 0.2 ($\tau_{med}$: 3.3 ps) |
| 0.01 | 0.09 ($\tau_{max}$: 7 ps) *Loc: $C_8$* | 60.78 ($\tau_{min}$: 10.8 fs) *Loc: $G_{16}$* | 0.96 ($\tau_{med}$: 685 fs) |
| 0.025 | 0.44 ($\tau_{max}$: 1.5 ps) *Loc: $C_2$* | 94 ($\tau_{min}$: 7 fs) *Loc: $G_{14}$* | 5.2 ($\tau_{med}$: 130.9 fs) |
| 0.05 | 0.82 ($\tau_{max}$: 800 fs) *Loc: $C_2$* | 133 ($\tau_{min}$: 5 fs) *Loc: $G_{14}$* | 60.6 ($\tau_{med}$: 37 fs) |
| 0.1 | 1.68 ($\tau_{max}$: 392 fs) *Loc: $C_2$* | 205 ($\tau_{min}$: 3.2 fs) *Loc: $G_{14}$* | 30.9 ($\tau_{med}$: 22.4 fs) |

From the energy-time uncertainty principle, we can relate dwell time ($\tau_{dwell}$) to the total DOS broadening. By calculating the DOS of the DNA without decoherence and only including the contacts' self-energies, we can estimate $\tau_{dwell} = \hbar/FWHM_{DOS}$, where $FWHM_{DOS}$ is the full-width at half maximum of the DOS at the HOMO. The dwell time is qualitatively representative of the time an electron placed in the resonant energy level stays in the DNA before going out to the contacts in the phase coherent limit [42]. We calculated a $\tau_{dwell}$ ~11ps at the $HOMO$ energy for $\Gamma_{L(R)} = 0.6$ eV. The computed $\tau_{dwell}$ is much larger than $\hbar/(\Gamma_L + \Gamma_R) \sim 0.55$ fs, which is the leakage rate through the contact. This indicates that the 11 ps long dwell time arises from barriers intrinsic to the DNA itself. We note that Conron *et al.* [43] reported a hopping time of approximately 232 ps between guanines in a G-stack of a different sequence, which is more than an order of magnitude larger than $\tau_{dwell}$. Their model is based on hopping transport, which allows for relaxation of the electron's energy. Depending on relative timescales of energy relaxation and charge transport through DNA, the coherent model may predict either faster or slower electron transfer, highlighting an issue that warrants further investigation.

We calculate the coherence lifetime ($\tau_{coh} = \hbar/\max(\Gamma_m)$) at the HOMO, which is the amount of time an electron residing at an energy level retains its quantum phase information. We see from Table 2 that even for a small $D_o = 0.001$ eV$^2$, $\tau_{coh} = \tau_{min} = 33$ fs is 300 times smaller than the dwell time (11 ps). As $D_o$ increases, the coherence lifetime becomes more than 3 orders of magnitude smaller than $\tau_{dwell}$. That is, dephasing events frequently occur while the electron is residing at the HOMO before it leaves the DNA. The calculated coherence lifetime is within the femtosecond-picosecond range reported in literature for solvated aromatic molecules [44–46], which translates to decoherence strengths of 1-130 meV, the $D_o$ values we choose to give decoherence rates in this range. The relative values of coherence and dwell time indicate that phase coherent transport across the entire length of this particular nine base pair long sequence will not occur.

# 5 Partitioning and Spurious Eigenstates in the HOMO-LUMO Gap

To apply decoherence, we need to partition the system into blocks and apply the decoherence probes to each block. The physical meaning of this partitioning is how we perceive the electron movement throughout the DNA and the locations at which an electron interacts with the environment and loses coherence. Atomic partitioning is the case where we assume that the electron enters the fictitious decoherence probe at each atom. As a result, inter-orbital mixing due to decoherence at the atomic location is allowed. This is a conservative approach. Another method of partitioning the system is by chemical intuition; we know that DNA consists of a chain of weakly coupled bases. These bases along with their sugar and backbone consist of aromatic rings that contain energetically unique molecular orbitals responsible for charge transport. Therefore, distribution of the charge density on the atoms depends on the chemical group they belong to [47]. In addition, there is a weak coupling between the neighboring bases (100 meV or less [48]). Therefore, nucleotide partitioning can be considered a natural scheme where the electron may reside on the localized molecular orbitals (LMOs) of a nucleotide and loses coherence. In this scenario, the decoherence probe self-energy is applied to the LMOs of each nucleotide, introducing energy shifts via its real part and level broadening through its imaginary part. Although, in theory, any arbitrary partitioning scheme can be made, one needs to be careful not to cause *artificial transport paths* between the nucleotides, which leads to unphysical shortcuts to the travel path of the electron as it traverses the DNA (discussed further in Section 5.1).

Partitioning the system into nucleotides or base pairs helps create a more meaningful picture of how the density of states (or molecular orbitals) is distributed along the DNA molecules. Each isolated partition will consist of its localized molecular orbitals (see Figure 10, **1-3**). Upon partitioning, we use a unitary transformation to transform the system Hamiltonian consisting of atomic orbitals into a diagonalized Hamiltonian (equation (7)). The diagonal elements of each block now represent the LMOs *of each partition*. If we further take all these partitions as one big block by including the orange interaction in Fig 12-3, we will find the orbitals of the whole DNA molecule (i.e., the *system's energy levels, the eigenvalue(H)*) (Figure 10, **4**). Grouping all atoms together as in step **4** will cause unphysical transport paths, which we discuss in the following section.

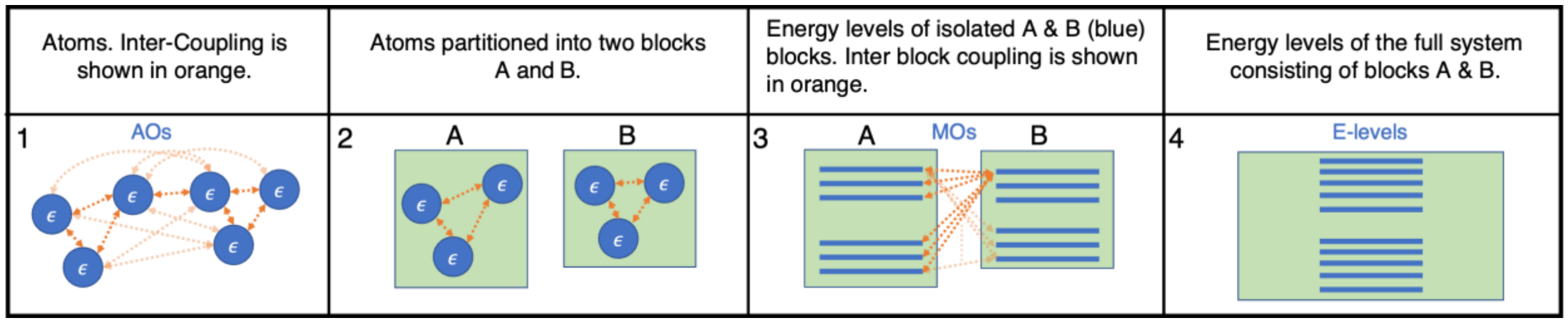

Figure 10 The system point of view before and after partitioning. **1)** The initial system is viewed as atoms with their localized atomic orbitals (AOs) interacting with each other. **2)** The partitioning starts by choosing the atoms that we want to group together. **3)** After applying the unitary transformation ($\mathrm{H} = U^{\dagger}H_0U$), we view the system as two interacting blocks with their localized molecular orbitals (LMOs) instead of atomic orbitals. **4)** If we take all partitions as one block, we get the energy levels of the system, however, we lose the spatial information of these energy levels.

Partitioning of DNA to add decoherence probes in reference [36] created two issues that requires attention. First, the scattering rates had a peak at the various energy levels $\tilde{\epsilon}_{k,m}$ of equation (14) which depended on the partitioning scheme. A better model would however make the scattering rates immune to the partitioning scheme adopted. Second, partitioning created "dangling bonds", which resulted in spurious eigen values (those that do not exist in the DNA) where the DOS and/or Transmissions have unphysical peaks. The model in this paper resolves both issues.

One key element discussed in the appendix of [36] is that partitioning DNA leads to breaking covalent C-O bonds between the sugar group at the backbone of one nucleotide and the phosphate group of its neighboring nucleotide, causing dangling bonds to appear as spurious eigenstates in the HOMO-LUMO gap. Hence, with the *E-dep* model, we can have transmission increase appearing deep in the energy gap region. In ref [36], we tried to address this drawback by not applying decoherence probes to spurious eigenstates that appear in the HOMO-LUMO gap. On the other hand, we can systematically overcome this issue using the *DOS-weighted* decoherence model in this manuscript. The transmission plot in Figure 11 goes from HOMO region to the HOMO-LUMO gap of 3'-CCCGCGCCC-5'. We notice that the *E-dep* model displays increase in transmission in the HOMO-LUMO gap region due to the spurious eigenstates resulting from the nucleotide partitioning scheme. However, the *DOS-weighted* decoherence model does not show much increase in transmission in the HOMO-LUMO gap even for large $D_o = 0.1$ eV$^2$. The reason is that the scattering rate in this model relies on the localized DOS of the full DNA system with contacts via eqt (25). Therefore, the maximum broadening is reached where the localized DOS is high at each partition $m$, which happens at energies equal to the eigenstates of the overall system. Therefore, large decoherence or transmission values do not appear deep in the HOMO-LUMO gap even after partitioning the system.

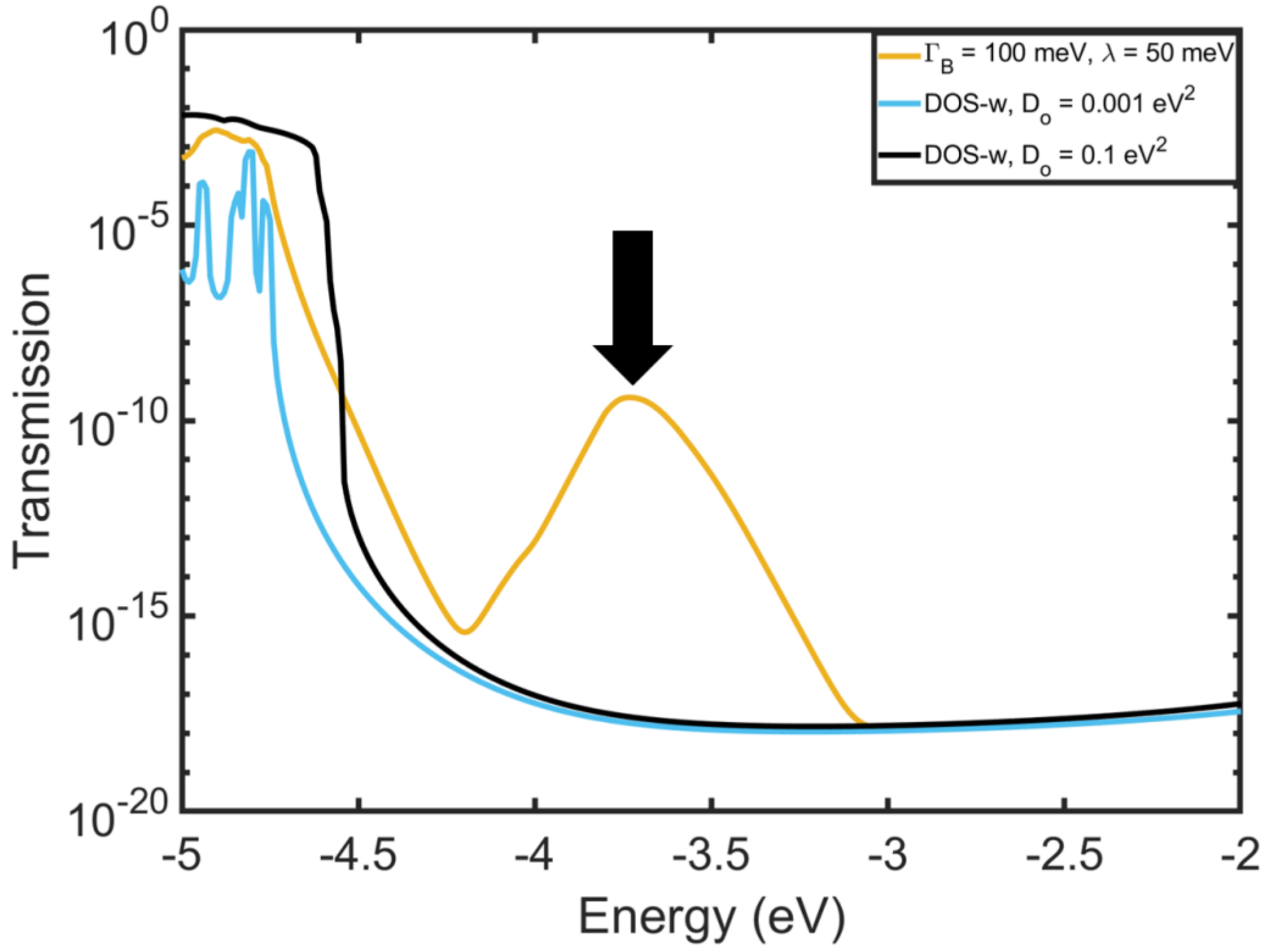

Figure 11 Transmission plot displaying E-dep decoherence transmission increase deep in the HOMO-LUMO gap due to spurious peaks, the DOS-weighted decoherence probe model helps resolve this issue.

## 5.1 Partitioning Scheme Comparison

Theoretically, the decoherence probes can be applied to any desired partitioning scheme. However, carelessly grouping several nucleotides under one partition can cause artificial tunneling shortcuts for the decoherent transmission. In this section, we compare different partitioning schemes: i) atomic partitioning, ii) nucleotide partitioning, and iii) the extreme case of a 3-block partitioning scheme. The atomic partitioning means that every atom is assigned a decoherence probe, whereas nucleotide partitioning results in 18 blocks as defined in Section 2.1. Figure 12 right, shows the 3-block partitioning, which consists of two contact blocks (yellow) having a nucleotide each and a large block for the remaining nucleotides (blue). We calculated the transmission using the *DOS-weighted* decoherence model with $D_o = [0.001, 0.01, 0.1]$ eV$^2$, $\Gamma_{L(R)} = 600$ meV, and $\eta_{Tr} = \eta_{DOS} = 0.1\%$. When comparing the partitioning schemes (Figure 12 left), the transmission amplitude increases as the number of partitions decreases.

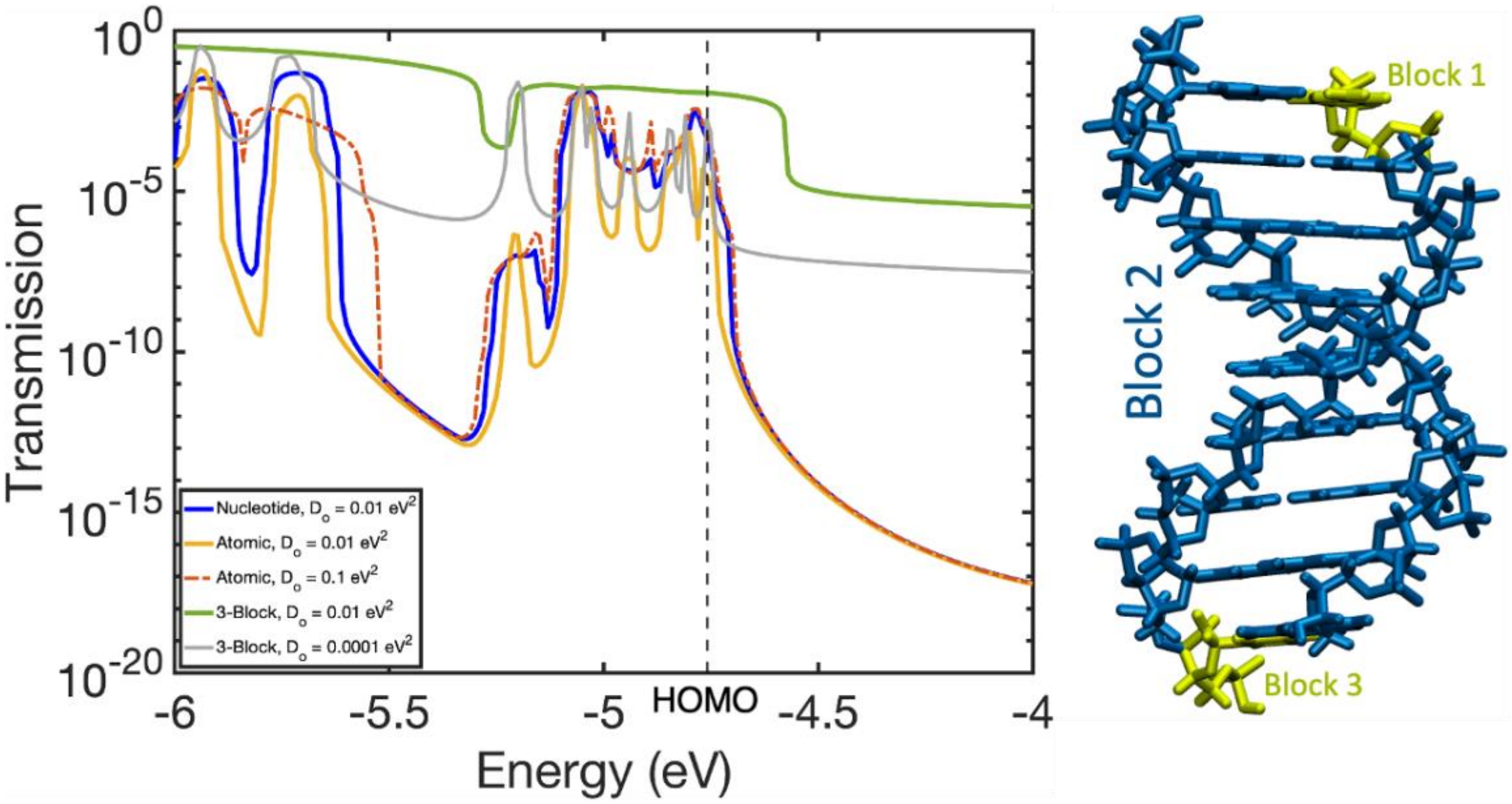

Figure 12 **Left** Transmission plot displaying the comparison between the partitioning schemes. **Right** 3-Block partitioning scheme of the DNA, the two yellow colors represent two blocks (for the contacts), the blue region represents the block attached to a single decoherence probe.

Concerning the model formalism, the more we partition the system into blocks, the lower the local-DOS value will be per block $m$ (the summation term in equation (26)). Therefore, by fixing $D_o$ and employing atomic partitioning, we are reducing the maximum decoherence value at any given block $m$ . In contrast, reducing the number of partitions means grouping more atoms together into one block, and their DOS contribution is summed together, increasing in the scattering rate for a fixed $D_o$. To elaborate, the transmission results in Figure 12 show us that setting $D_o = 0.1$ eV$^2$ with atomic partitioning (red dashed curve) can be almost equal to the nucleotide partitioning with $D_o = 0.01$ eV$^2$ (blue curve) for E > -5.5 eV. As for the unrealistic increase in transmission amplitude of the 3-block partitioning scheme (green curve in Figure 12), it can be attributed to the unphysical impact of using low number of blocks, as illustrated by the example in Figure 13. The extreme case of having most bases connected to one decoherence probe (blue partition in Figure 13) would allow a scenario where the electron scatters from the first base into the decoherence probe, only to be injected back into the system at the end base, causing an unphysical shortcut in transmission. We note that only for very low $D_o = 0.0001\ \mathrm{eV}^2$ (grey curve in Figure 12) does the transmission resemble the other cases, though it similarly overestimates off-resonant transmission (E~-5.5 eV) as in the higher $D_o$ case (green curve). We emphasize that the decoherence probes inject and extract the scattering electron at the same probe site $m$. Therefore, if we partition the DNA into one large segment, the decoherence probe only sees localized orbitals of this large segment (blue region in Figure 15). Hence, careful partitioning of the system is required to not cause unphysical transport paths for the electron.

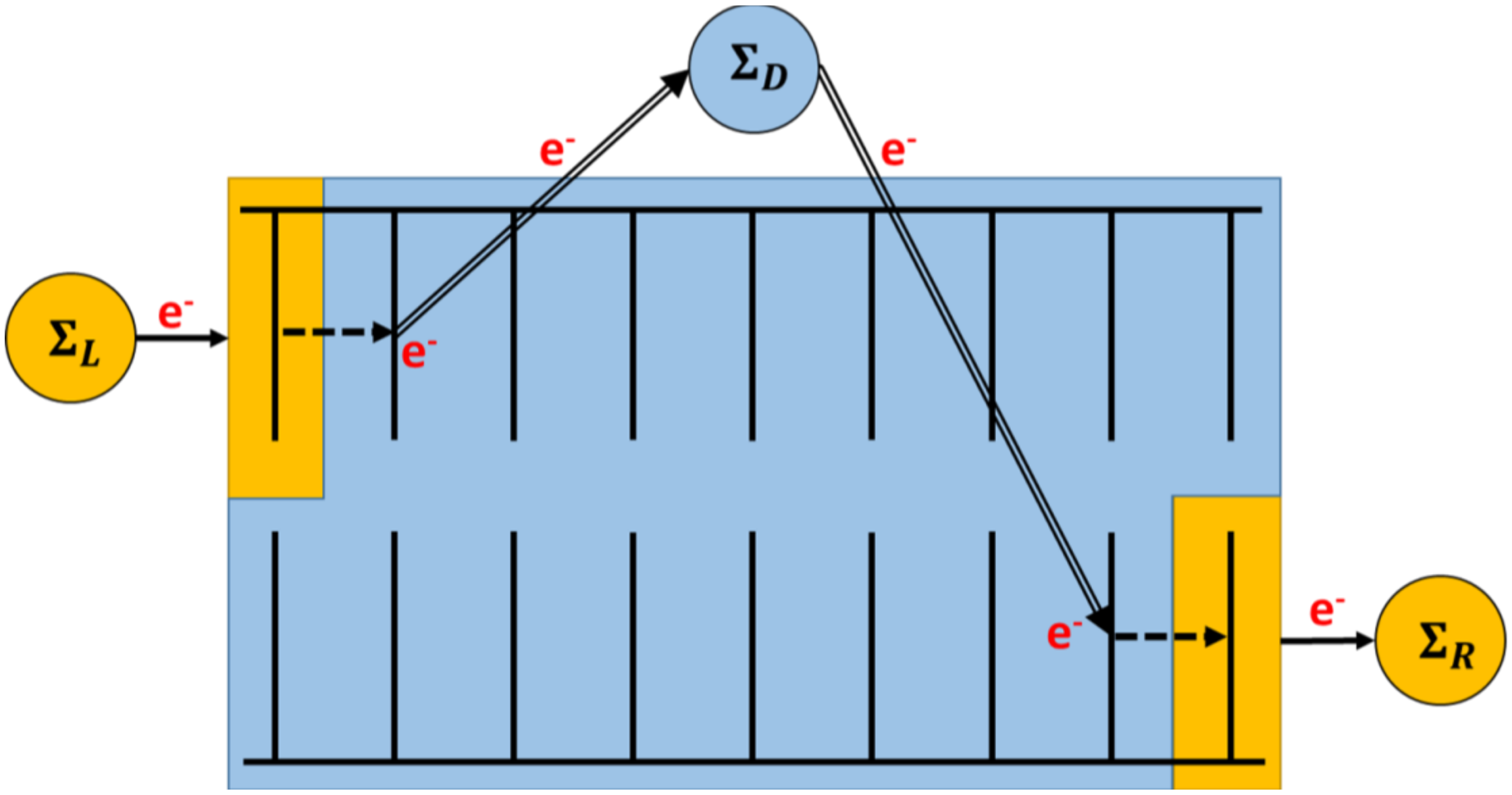

Figure 13 Illustrative example of how low number of partitions (3-blocks for instance) in a ds-DNA creates unphysical shortcuts in charge transport. The blue block includes several nucleotides coupled with one decoherence probe. The blue partition now has the LMOs of all nucleotides within it and the spatial information of these orbitals is no longer apparent to the decoherence probe. Therefore, the electron scattered into the probe could be re-injected into the system anywhere within the block. This result causes higher transmission but is considered unphysical since the electron is taking a shortcut and skipping over several molecules.

# 6 Conclusion

Molecular structures such as DNA present promising opportunities for nanoelectronic devices and sensing applications. In these systems, the small physical dimensions make atomic-level details critical to charge transport behavior, necessitating atomistic simulations with quantum mechanical treatment. We employ a combination of density functional theory and Green's function-based quantum charge transport formalism to model charge propagation through DNA. Modeling phase-breaking scattering events in such systems is particularly challenging due to limited experimental data and the large number of influencing factors, including solvent effects, room temperature operation, and the intrinsic structural fluctuations of the molecule. Several strategies have been proposed for modeling decoherent charge transport in nanoscale DNA. Among them, the decoherence probes method stands out for its balance between computational efficiency and accuracy. In this approach, phenomenological probes are attached to individual atoms or a defined group of atoms to simulate electron scattering and phase loss at those locations. Previous implementations of the decoherence probe model have faced notable limitations. An energy-independent scattering rate led to excessive broadening of energy levels and an unphysical increase in the density of states within the HOMO-LUMO gap. Introducing an energy-dependent scattering rate addressed this issue but introduced spurious peaks within energy gaps and required an additional fitting parameter. To overcome these drawbacks, we redefine the scattering rate to be directly proportional to the DOS, yielding a spatially and energy-dependent rate that eliminates the spurious peaks. The model iteratively solves for the decoherence self-energy and DOS until convergence, with the final scattering rates and coherence lifetimes reported for the DNA system.

We also examine the implications of partitioning of the system for applying decoherence probes. Dividing the system into groups of atoms per block can artificially break covalent bonds, introducing spurious peaks in the HOMO-LUMO and other energy gaps. The proposed DOS-weighted model mitigates this issue by linking the scattering rate to the DOS of the entire system rather than to the energy levels of partitions which produce spurious energy levels and transmission peaks. Furthermore, we emphasize the importance of carefully choosing the partitioning scheme, as improper grouping can create unphysical shortcuts in the transport path, leading to an overestimation of transmission.

# 7 References

[1] P. W. K. Rothemund, *Folding DNA to Create Nanoscale Shapes and Patterns*, Nature 440, 297 (2006).

[2] Y. Ke, L. L. Ong, W. M. Shih, and P. Yin, *Three-Dimensional Structures Self-Assembled from DNA Bricks*, Science (1979) 338, 1177 (2012).

[3] J. M. Artés, Y. Li, J. Qi, M. P. Anantram, and J. Hihath, *Conformational Gating of DNA Conductance*, Nat Commun 6, 8870 (2015).

[4] Y. Li, J. M. Artés, J. Qi, I. A. Morelan, P. Feldstein, M. P. Anantram, and J. Hihath, *Comparing Charge Transport in Oligonucleotides: RNA:DNA Hybrids and DNA Duplexes*, J Phys Chem Lett 7, 1888 (2016).

[5] T. Harashima, C. Kojima, S. Fujii, M. Kiguchi, and T. Nishino, *Single-Molecule Conductance of DNA Gated and Ungated by DNA-Binding Molecules*, Chemical Communications 53, 10378 (2017).

[6] C. Guo, K. Wang, E. Zerah-Harush, J. Hamill, B. Wang, Y. Dubi, and B. Xu, *Molecular Rectifier Composed of DNA with High Rectification Ratio Enabled by Intercalation*, Nat Chem 8, 484 (2016).

[7] L. Xiang, J. L. Palma, Y. Li, V. Mujica, M. A. Ratner, and N. Tao, *Gate-Controlled Conductance Switching in DNA*, Nat Commun 8, 1 (2017).

[8] H. Mohammad, B. Demir, C. Akin, B. Luan, J. Hihath, E. E. Oren, and M. P. Anantram, *Role of Intercalation in the Electrical Properties of Nucleic Acids for Use in Molecular Electronics*, Nanoscale Horiz 6, 651 (2021).

[9] C. Liu, L. Guo, B. Zhang, and L. Lu, *Graphene Quantum Dots Mediated Electron Transfer in DNA Base Pairs*, RSC Adv 9, 31636 (2019).

[10] J. Hihath, S. Guo, P. Zhang, and N. Tao, *Effects of Cytosine Methylation on DNA Charge Transport*, Journal of Physics: Condensed Matter 24, 164204 (2012).

[11] J. Qi, N. Govind, and M. P. Anantram, *The Role of Cytosine Methylation on Charge Transport through a DNA Strand*, J Chem Phys 143, 094306 (2015).

[12] Y. Li, J. M. Artés, B. Demir, S. Gokce, H. M. Mohammad, M. Alangari, M. P. Anantram, E. E. Oren, and J. Hihath, *Detection and Identification of Genetic Material via Single-Molecule Conductance*, Nat Nanotechnol 13, 1167 (2018).

[13] A. De, B. Lu, Y. P. Ohayon, K. Woloszyn, W. Livernois, L. Perren, C. fan Yang, C. Mao, A. S. Botana, J. Hihath, et al., *Transmetalation for DNA-Based Molecular Electronics*, Small 21, 2411518 (2025).

[14] H. Mohammad, L. Alsaleh, A. Alotaibi, O. Alolaiyan, T. Takahashi, M. P. Anantram, and T. Nishino, *DNA Conductance Modulation* via *Aptamer Binding*, Nanoscale 17, 8035 (2025).

[15] C. Adessi, S. Walch, and M. P. Anantram, *Environment and Structure Influence on DNA Conduction*, Phys Rev B Condens Matter Mater Phys 67, 1 (2003).

[16] S. R. Patil, H. Mohammad, V. Chawda, N. Sinha, R. K. Singh, J. Qi, and M. P. Anantram, *Quantum Transport in DNA Heterostructures: Implications for Nanoelectronics*, ACS Appl Nano Mater 4, 10029 (2021).

[17] A. De, H. Mohammad, Y. Wang, R. Kubendran, A. K. Das, and M. P. Anantram, *Performance Analysis of DNA Crossbar Arrays for High-Density Memory Storage Applications*, Sci Rep 13, 6650 (2023).

[18] P. T. Bui, T. Nishino, H. Shiigi, and T. Nagaoka, *One-by-One Single-Molecule Detection of Mutated Nucleobases by Monitoring Tunneling Current Using a DNA Tip*, Chemical Communications 51, 1666 (2015).

[19] S. Afsari, L. E. Korshoj, G. R. Abel, S. Khan, A. Chatterjee, and P. Nagpal, *Quantum Point Contact Single-Nucleotide Conductance for DNA and RNA Sequence Identification*, ACS Nano 11, 11169 (2017).

[20] Y. Li, L. Xiang, J. L. Palma, Y. Asai, and N. Tao, *Thermoelectric Effect and Its Dependence on Molecular Length and Sequence in Single DNA Molecules*, Nat Commun 7, 1 (2016).

[21] D. Porath, A. Bezryadin, S. De Vries, and C. Dekker, *Direct Measurement of Electrical Transport through DNA Molecules*, Nature 2000 403:6770 403, 635 (2000).

[22] A. Rakitin, P. Aich, C. Papadopoulos, Y. Kobzar, A. S. Vedeneev, J. S. Lee, and J. M. Xu, *Metallic Conduction through Engineered DNA: DNA Nanoelectronic Building Blocks*, Phys Rev Lett 86, 3670 (2001).

[23] C. Nogues, S. R. Cohen, S. Daube, N. Apter, and R. Naaman, *Sequence Dependence of Charge Transport Properties of DNA*, J Phys Chem B 110, 8910 (2006).

[24] H. Cohen, C. Nogues, R. Naaman, and D. Porath, *Direct Measurement of Electrical Transport through Single DNA Molecules of Complex Sequence*, Proceedings of the National Academy of Sciences 102, 11589 (2005).

[25] Y. Wang, B. Demir, H. Mohammad, E. E. Oren, and M. P. Anantram, *Computational Study of the Role of Counterions and Solvent Dielectric in Determining the Conductance of B-DNA*, Phys Rev E 107, 044404 (2023).

[26] W. N. Zaharim, S. Sulaiman, L. S. Ang, S. Shamsuddin, N. Najimudin, G. Azzam, and I. Watanabe, *DFT Investigation of the Electronic Structure, Non-Covalent Interactions, and Muon Hyperfine Interactions in Short Adenine–Thymine Double-Strand DNA*, ACS Omega 10, 30670 (2025).

[27] B. Demir, H. Mohammad, M. P. Anantram, and E. E. Oren, *DNA–Au (111) Interactions and Transverse Charge Transport Properties for DNA-Based Electronic Devices*, Physical Chemistry Chemical Physics 25, 16570 (2023).

[28] P. Karasch, D. A. Ryndyk, and T. Frauenheim, *Vibronic Dephasing Model for Coherent-to-Incoherent Crossover in DNA*, Phys Rev B 97, 195401 (2018).

[29] R. Gutiérrez, S. Mandal, and G. Cuniberti, *Quantum Transport through a DNA Wire in a Dissipative Environment*, Nano Lett 5, 1093 (2005).

[30] R. Gutiérrez, S. Mandal, and G. Cuniberti, *Dissipative Effects in the Electronic Transport through DNA Molecular Wires*, Phys Rev B 71, 235116 (2005).

[31] T. Kubař, R. Gutiérrez, U. Kleinekathöfer, G. Cuniberti, and M. Elstner, *Modeling Charge Transport in DNA Using Multi-Scale Methods*, Phys Status Solidi B Basic Res 250, 2277 (2013).

[32] H. L. Engquist and P. W. Anderson, *Definition and Measurement of the Electrical and Thermal Resistances*, Phys Rev B 24, 1151 (1981).

[33] M. Büttiker, Y. Imry, R. Landauer, and S. Pinhas, *Generalized Many-Channel Conductance Formula with Application to Small Rings*, Phys Rev B 31, 6207 (1985).

[34] M. Buttiker, *Coherent and Sequential Tunneling in Series Barriers*, IBM J Res Dev 32, 63 (1988).

[35] J. Qi, N. Edirisinghe, M. G. Rabbani, and M. P. Anantram, *Unified Model for Conductance through DNA with the Landauer-Büttiker Formalism*, Phys Rev B Condens Matter Mater Phys 87, 1 (2013).

[36] H. Mohammad and M. P. Anantram, *Charge Transport through DNA with Energy-Dependent Decoherence*, Phys Rev E 108, 044403 (2023).

[37] A. Svizhenko and M. P. Anantram, *Effect of Scattering and Contacts on Current and Electrostatics in Carbon Nanotubes*, Phys Rev B Condens Matter Mater Phys 72, (2005).

[38] D. A. Case, R. M. Betz, D. S. Cerutti, C. I. T.E., T. A. Darden, R. E. Duke, T. J. Giese, H. Gohlke, A. W. Goetz, N. Homeyer, et al., *Amber 2020*, University of California, San Francisco (2020).

[39] D. J. Frisch, M. J.; Trucks, G. W.; Schlegel, H. B.; Scuseria, G. E.; Robb, M. A.; Cheeseman, J. R.; Scalmani, G.; Barone, V.; Petersson, G. A.; Nakatsuji, H.; Li, X.; Caricato, M.; Marenich, A. V.; Bloino, J.; Janesko, B. G.; Gomperts, R.; Mennucci, B.; Hratch, *Gaussian 16, Revision C.01*.

[40] Y. Wang, B. Demir, H. Mohammad, E. E. Oren, and M. P. Anantram, *A Computational Study of the Role of Counterions and Solvent Dielectric in Determining the Conductance of B-DNA*, BioRxiv 2023.03.29.534812 (2023).

[41] J. L. D'Amato and H. M. Pastawski, *Conductance of a Disordered Linear Chain Including Inelastic Scattering Events*, Phys Rev B 41, 7411 (1990).

[42] B. Küng, C. Rössler, M. Beck, J. Faist, T. Ihn, and K. Ensslin, *Quantum Dot Occupation and Electron Dwell Time in the Cotunneling Regime*, New J Phys 14, 083003 (2012).

[43] S. M. M. Conron, A. K. Thazhathveetil, M. R. Wasielewski, A. L. Burin, and F. D. Lewis, *Direct Measurement of the Dynamics of Hole Hopping in Extended DNA G-Tracts. An Unbiased Random Walk*, J Am Chem Soc 132, 14388 (2010).

[44] W. W. Parson, *Temperature Dependence of the Rate of Intramolecular Electron Transfer*, Journal of Physical Chemistry B 122, 8824 (2018).

[45] W. W. Parson, *Vibrational Relaxations and Dephasing in Electron-Transfer Reactions*, J Phys Chem B 120, 11412 (2016).

[46] W. W. Parson, *Effects of Free Energy and Solvent on Rates of Intramolecular Electron Transfer in Organic Radical Anions*, (2017).

[47] I. V. Kurnikov and D. N. Beratan, *Ab Initio Based Effective Hamiltonians for Long-range Electron Transfer: Hartree–Fock Analysis*, J Chem Phys 105, 9561 (1996).

[48] H. Mehrez and M. P. Anantram, *Interbase Electronic Coupling for Transport through DNA*, Phys Rev B Condens Matter Mater Phys 71, (2005).